\documentclass[aps,prfluids,reprint,superscriptaddress,showpacs,preprintnumbers,onecolumn,floatfix,longbibliography]{revtex4-2}
\usepackage{hyperref}
\usepackage{amsmath,latexsym}
\usepackage{graphicx}
\usepackage{siunitx}
\usepackage{textcomp}
\usepackage{wasysym}
\usepackage{amssymb}
\usepackage{MnSymbol}
\usepackage{oplotsymbl}
\usepackage{setspace}
\usepackage{svg}
\usepackage{mathtools}
\usepackage{wrapfig}
\usepackage{rotating}
\usepackage{epstopdf}
\usepackage{setspace}
\makeatletter
\let\@currsize\normalsize
\makeatother
\usepackage{etoolbox}
\usepackage{tikz}
\usepackage{multirow}
\usepackage{tablefootnote}
\usepackage{booktabs, makecell}

\usepackage{threeparttable}
\usepackage{setspace}
\usepackage{svg}

\definecolor{water}{rgb}{0.5,0.5,0.5}
\definecolor{2M50ppm}{rgb}{1,0,0}
\definecolor{2M100ppm}{rgb}{1,0.285714285714286,0}
\definecolor{2M200ppm}{rgb}{1,0.571428571428571,0}
\definecolor{2M300ppm}{rgb}{1,0.857142857142857,0}
\definecolor{2M400ppm}{rgb}{1,1,1}
\definecolor{2M500ppm}{rgb}{1,1,1}
\definecolor{4M25ppm}{rgb}{0,0,1}
\definecolor{4M50ppm}{rgb}{0,0.181818181818182,0.909090909090909}
\definecolor{4M100ppm}{rgb}{0,0.363636363636364,0.818181818181818}
\definecolor{4M200ppm}{rgb}{0,0.545454545454545,0.727272727272727}
\definecolor{4M250ppm}{rgb}{0,0.727272727272727,0.636363636363636}
\definecolor{4M300ppm}{rgb}{0,0.909090909090909,0.545454545454545}
\definecolor{8M12ppm}{rgb}{0,0.500000000000000,0.400000000000000}
\definecolor{8M50ppm}{rgb}{0.222222222222222,0.611111111111111,0.400000000000000}
\definecolor{8M75ppm}{rgb}{0.444444444444444,0.722222222222222,0.400000000000000}
\definecolor{8M100ppm}{rgb}{0.666666666666667,0.833333333333333,0.400000000000000}
\definecolor{8M150ppm}{rgb}{0.888888888888889,0.944444444444444,0.400000000000000}

\definecolor{laminar}{rgb}{0,0.6,0.3}
\definecolor{puff}{rgb}{0.9290,0.6940,0.1250}
\definecolor{EIS}{rgb}{0,0.5,1}
\definecolor{EIT}{rgb}{0.75,0,0.75}
\definecolor{NT}{rgb}{1,0.2,0.2}

\newrobustcmd*{\mysquare}[1]{\tikz{\filldraw[draw=#1,fill=#1] (0,0)
rectangle (0.2cm,0.2cm);}}

\newrobustcmd*{\mycircle}[1]{\tikz{\filldraw[draw=#1,fill=#1] (0,0) circle [radius=0.1cm];}}

\newrobustcmd*{\mytriangle}[1]{\tikz{\filldraw[draw=#1,fill=#1] (0,0) --
(0cm,0.2cm) -- (0.2cm,0.1cm);}}

\newcommand{\sqdiamond}[1][fill=black]{\tikz [x=1.2ex,y=1.85ex,line width=.3ex,line join=round, yshift=-0.285ex] \draw  [#1]  (0,.5) -- (.5,1) -- (1,.5) -- (.5,0) -- (0,.5) -- cycle;}%

\newcommand{\MyDiamond}[1][fill=black]{\mathop{\raisebox{-0.275ex}{$\sqdiamond[#1]$}}}

\usepackage{todonotes}
\usepackage{comment}
\allowdisplaybreaks[1]
\begin{document}
\title{Spatio-temporal Signatures of Elasto-inertial Turbulence in Viscoelastic Planar Jets}
\date{\today}
\author{Sami Yamani}
\email[Corresponding author. ]{syamani@mit.edu}
\affiliation{Hatsopoulos Microfluids Laboratory, Department of Mechanical Engineering, Massachusetts Institute of Technology, Cambridge, Massachusetts 02139, USA}
\author{Yashasvi Raj}
\affiliation{Hatsopoulos Microfluids Laboratory, Department of Mechanical Engineering, Massachusetts Institute of Technology, Cambridge, Massachusetts 02139, USA}
\author{Tamer~A.~Zaki}
\affiliation{Department of Mechanical Engineering, Johns Hopkins University, Baltimore, Maryland 21218, USA}
\author{Gareth H. McKinley}
\affiliation{Hatsopoulos Microfluids Laboratory, Department of Mechanical Engineering, Massachusetts Institute of Technology, Cambridge, Massachusetts 02139, USA}
\author{Irmgard Bischofberger}
\email[Corresponding author. ]{irmgard@mit.edu}
\affiliation{Hatsopoulos Microfluids Laboratory, Department of Mechanical Engineering, Massachusetts Institute of Technology, Cambridge, Massachusetts 02139, USA}

\begin{abstract}

\noindent The interplay between viscoelasticity and inertia in dilute polymer solutions at high deformation rates can result in inertio-elastic instabilities. The nonlinear evolution of these instabilities generates a state of turbulence with significantly different spatio-temporal features compared to Newtonian turbulence, termed elasto-inertial turbulence (EIT). We explore EIT by studying the dynamics of a submerged planar jet of a dilute aqueous polymer solution injected into a quiescent tank of water using a combination of schlieren imaging and laser Doppler velocimetry (LDV). We show how fluid elasticity has a nonmonotonic effect on the jet stability depending on its magnitude, creating two distinct regimes in which elastic effects can either destabilize or stabilize the jet. In agreement with linear stability analyses of viscoelastic jets, an inertio-elastic shear-layer instability emerges near the edge of the jet for small levels of elasticity, independent of bulk undulations in the fluid column. The growth of this disturbance mode destabilizes the flow, resulting in a turbulence transition at lower Reynolds numbers and closer to the nozzle compared to the conditions required for the transition to turbulence in a Newtonian jet. Increasing the fluid elasticity merges the shear-layer instability into a bulk instability of the jet column. In this regime, elastic tensile stresses generated in the shear layer act as an ``elastic membrane'' that partially stabilizes the flow, retarding the transition to turbulence to higher levels of inertia and greater distances from the nozzle. In the fully turbulent state far from the nozzle, planar viscoelastic jets exhibit unique spatio-temporal features associated with EIT. The time-averaged angle of jet spreading, an Eulerian measure of the degree of entrainment, and the centerline velocity of the jets both evolve self-similarly with distance from the nozzle. The autocovariance of the schlieren images in the fully turbulent region of the jets shows coherent structures that are elongated in the streamwise direction, consistent with the suppression of streamwise vortices by elastic stresses. These coherent structures give a higher spectral energy to small frequency modes in EIT characterized by LDV measurements of the velocity fluctuations at the jet centerline. Finally, our LDV measurements reveal a frequency spectrum characterized by a –3 power-law exponent, different from the well-known –5/3 power-law exponent characteristic of Newtonian turbulence.

% LDV measurements of the velocity fluctuations at the jet centerline reveal a frequency spectrum characterized by a –3 power-law exponent, different from the well-known –5/3 power-law exponent characteristic of Newtonian turbulence. We show that the higher spectral energy of long wavelength modes in the EIT state results in coherent structures that are elongated in the streamwise direction, consistent with the suppression of streamwise vortices by elastic stresses. 
\end{abstract}

\maketitle
\section{Introduction}

% \noindent \textcolor{red}{Please note that marked in red are: 1. Reference to appendix material that I have not added to appendix yet, 2. A few parts that we do not have a good physical explanations for them.}

\noindent Dilute polymer solutions offer the intriguing engineering opportunity of passively reducing frictional drag in turbulent flows in pipelines \cite{Burger1982}, in boundary layers around the body of marine vehicles \cite{marc2014mitigation}, or in open channel irrigation flows \cite{bouchenafa2021water}. In turbulent pipe flows, the addition of minute quantities of a long chain polymer causes a reduction in the frictional pressure drop per unit length by as much as $40\%$ \cite{toms1948some, mysels1971early}. The injection of a concentrated polymer solution into the near-wall region of a turbulent boundary layer can reduce skin friction resistance by up to $75\%$, which can potentially translate into substantial savings in propulsive power requirements and fuel costs \cite{winkel2009high}.
% Experiments conducted in towing tanks and  water tunnels have shown that
As a consequence of a maximum drag reduction (MDR) asymptote that is independent of polymer concentration, one cannot indefinitely continue to reduce drag by increasing the polymer concentration \cite{Virk1967,Virk1975}. While MDR has long been accepted to be an asymptotic limit for drag reduction, it was shown recently that the spatio-temporal dynamics of the flow continue to evolve with increasing polymer concentration \cite{zhu2021nonasymptotic}, and a new state of turbulence driven by viscoelasticity, termed elasto-inertial turbulence (EIT), emerges \cite{Samanta2013}. EIT encompasses a broad range of Reynolds numbers (Re) \cite{choueiri2021experimental}, characterizing the ratio of inertial forces to viscous forces, 
 and --- for an appropriate choice of parameters --- drag reduction beyond the MDR limit can be achieved by passing through a relaminarized state \cite{Choueiri2018, lopez2019dynamics}. While significant progress has been made in understanding EIT for internal flows, limited work has addressed external flows such as free shear layers. %which are the basis for drag reduction in external flows. 
In this work, we investigate inertio-elastic instabilities and turbulence in a canonical geometry commonly used for studying free shear layers; a submerged planar jet of dilute aqueous polymer solution injected into a quiescent tank of water. 

\noindent A fully turbulent state develops from the growth of small-amplitude unstable modes and their nonlinear interactions. This turbulent state efficiently mixes the injected polymer solution into the background fluid. The driving mechanisms promoting instability have been investigated through temporal linear stability analysis using the Oldroyd-B model \cite{bird1987dynamicsv1} for the planar mixing layer geometry. An increase in the elasticity number (El), characterizing the ratio between the relaxation time scale of the polymer and the vorticity diffusion time scale of the fluid, stabilizes long wavelength instabilities due to an increase in the magnitude of the normal stress differences in the flow \cite{azaiez1994linear}. As discussed in E.\,J.\,Hinch's appendix of \cite{azaiez1994linear}, the stabilizing effect of normal stresses is the result of an effective elastic membrane at the shear layer interface, which generates effects akin to those of surface tension on the Kelvin-Helmholtz instability. It should be noted that this analysis was performed in the dual limit of high Reynolds number and high Weissenberg number (Wi), where Wi characterizes the ratio of the polymer relaxation time scale to the convective time scale of the flow. In this limit, the effects of momentum diffusion, stress relaxation during disturbance growth, and the role of polymer shear stresses are neglected and only the normal stress difference generated by the polymer is considered. A more recent spatio-temporal stability analysis \cite{ray2014absolute} studied the absolute instability of planar shear flow using both the Oldroyd-B and the finitely extensible nonlinear elastic (FENE-P) models \cite{bird1987dynamicsv1}. The stabilization of the instability observed in the Oldroyd-B model is shown to be an asymptotic limit that is only valid for large El, while the initial effects of increasing El are, in fact, destabilizing at low and moderate elasticity numbers \cite{ray2014absolute}. Moreover, the local shear rate distribution across the mixing layer plays an important role in determining whether elasticity has a stabilizing or destabilizing effect; only for high enough local shear rates are the normal stress differences large enough to stabilize the flow. A scaled Weissenberg number based on local shear rate and finite extensibility of the polymer chains ($L_{max}$) was introduced to characterize the stabilizing or destabilizing effect of elasticity, where $L_{max}$ denotes the ratio of the fully extended polymer chain length to the  root-mean-square end-to-end separation of the polymer in the equilibrium coiled state. Consistent with the stability analysis, two-dimensional (2D) \cite{azaiez1994numerical} and three-dimensional (3D) \cite{kumar1999direct} direct numerical simulations of this geometry report that the shear layer is more stable at large elasticity numbers and more unstable to 3D perturbations at small El, compared to a Newtonian shear layer. %In summary, increasing elasticity number at low and moderate values has a destabilizing effect, while a further increase in elasticity number has a stabilizing effect.  

% While linear stability analysis and DNS studies on undisturbed free shear flow are insightful, there are some challenges in experimentally studying undisturbed free shear flow. Thus, theoreticians have also studied 
% Linear stability analysis has further been done on submerged jets with various velocity profiles, a geometry used in several industrial applications \cite{Rallison1995, ray2015absolute}. 

\noindent In addition to the planar mixing layer, linear stability analyses have been also carried out on more complex flow geometries, including on submerged jets with different velocity profiles. To our knowledge, the earliest analysis on submerged jets is a temporal linear stability analysis on planar and axisymmetric jets with a parabolic velocity profile using the Oldroyd-B model \cite{Rallison1995}, again in the limit of high Reynolds number and high Weissenberg number. 
% The jet column, \textit{i.e.}, 
It was shown that the bulk fluid column of the jet can become unstable to sinuous or varicose modes of instability, and increasing the elasticity number stabilizes these instabilities through the same mechanism discussed by Hinch in the appendix of \cite{azaiez1994linear}. At small elasticity numbers, however, a local shear-layer mode of instability exists close to the edge of the jet with a higher wavenumber compared to the modes that destabilize the jet column. This shear-layer instability is independent of the jet geometry and the jet-column instability. At high elasticity numbers, computations show that this shear-layer mode merges with the jet-column mode \cite{Rallison1995}. Recent work on the absolute instability of viscoelastic planar jets using a spatio-temporal linear stability analysis \cite{ray2015absolute} introduces a shear layer thickness parameter that allows a more realistic jet velocity profile compared to the parabolic profile considered in \cite{Rallison1995} as well as a co- or counter-flow parameter allowing for flow in the surrounding fluid reservoir, and uses the FENE-P constitutive model to incorporate the impact of the finite extensibility of the polymer chains. This analysis also finds that increasing the elasticity number can have a destabilizing or stabilizing effect, depending on the velocity profile in the jet. 

\noindent The nonlinear evolution of instabilities in viscoelastic jets results in EIT, a state of turbulence that is significantly different from Newtonian turbulence. 
%While scientists and engineers have been working on the drag reduction application of viscoelastic flows for more than 70 years \cite{toms1948some, mysels1971early, Virk1967,Virk1975}, it was not until the past decade that scientists became interested in 
%how the presence of macromolecules modify the spatio-temporal features of turbulence.
Theoretical \cite{fouxon2003spectra}, computational \cite{dubief2013mechanism, thais2013spectral, watanabe2013hybrid,valente2014effect, valente2016energy}, and experimental \cite{vonlanthen2013grid, warholic1999influence,yamani2021spectral} work addressing how the presence of polymers modifies the spatio-temporal features of the turbulent state have suggested a turbulent kinetic energy spectrum that decays with a significantly steeper power-law for elasto-inertial turbulence when compared to the $-5/3$ power-law \cite{kolmogorov1941local} characterizing Newtonian turbulence. Polymers can temporarily store some of the fluctuating turbulent energy and release it back to the mean flow, inhibiting it from being transferred across the inertial range through the conventional Newtonian turbulent energy cascade \cite{balkovsky2000turbulent, balkovsky2001turbulence,valente2014effect,valente2016energy,abreu2022turbulent}. The higher wavenumbers of the turbulent spectrum thus have a lower power spectral density (PSD), resulting in a steeper power-law decay in EIT compared to Newtonian turbulence.

\noindent In wall-bounded flows, EIT consists of active and hibernating turbulent spots \cite{xi2010active, xi2010turbulent,Graham2014, pereira2017active, wang2017spatiotemporal} generated through transitory energy storage and release by the dissolved polymer chains in the fluid \cite{dubief2004coherent}. This energy exchange between macromolecules and the background flow results in alternating vortical and extensional regions \cite{dubief2013mechanism, Samanta2013} that partially or fully suppress the streamwise vortices \cite{li2007polymer, stone2004polymer, Samanta2013, Agarwal2014, sid2018two}. Computations \cite{Lee2017,Shekar2019, shekar2019self, shekar2021tollmien} show that one route to EIT is through nonlinearly amplified Tollmien-Schlichting waves that are self-sustained by viscoelasticity. In addition, computations \cite{garg2018viscoelastic, page2020exact,buza2022weakly,dubief2022first} and experiments \cite{choueiri2021experimental} reveal a second route in Poiseuille flow, in which a centerline instability distinguished by arrowhead-shaped coherent structures leads to EIT. This route is similar to that observed for the transition to purely elastic turbulence in the absence of inertia \cite{garg2018viscoelastic, page2020exact,buza2022weakly,dubief2022first,choueiri2021experimental}.

\noindent Transition to EIT in free shear layers has so far received relatively limited attention. %Early experimental studies added regular dye \cite{berman1985two} and fluoreceine \cite{gyr1996effects}  to visualize a jet of aqueous dilute polymer solution injected into a quiescent tank of water. In addition, a qualitative comparison used schlieren imaging to visualize the viscoelastic turbulent structures \cite{hibberd1982influence}. Velocity measurements were also carried out using pitot tubes \cite{white1967velocity}, piezoceramics \cite{shul1973measurement} and laser Doppler velocimetry (LDV) \cite{barker1973laser, vlasov1973average,berman1985two,berman1985two}. One common problem among all these studies is the lack of quantitative characterization of jet elasticity. Despite this problem, they show that the 
Viscoelastic jets have been reported to be more stable and to exhibit larger turbulent structures compared to Newtonian jets \cite{gyr1996effects, hibberd1982influence}, and the mean centerline velocity of a round jet follows the same self-similar scaling as a Newtonian jet \cite{white1967velocity}. Moreover, the viscoelastic jet has a smaller spreading angle and entrains less fluid compared to the Newtonian jet \cite{shul1973measurement, berman1985two, berman1986dispersion,gyr1996effects} due to smaller velocity fluctuations and hence a smaller turbulent kinetic energy (TKE) transfer rate~\cite{berman1986dispersion}. The turbulent kinetic energy spectrum of the viscoelastic jet calculated from LDV measurements of the velocity fluctuations exhibits a steeper decay compared to that of a Newtonian jet \cite{berman1985two,shul1973measurement}. A quantitative characterization of the role of jet elasticity though has so far been lacking.   %Hence, the viscoelastic spectrum is  higher than the Newtonian spectrum at low frequencies and lower than the Newtonian spectrum at higher frequencies \cite{berman1985two,shul1973measurement}. 
%A more recent theoretical and numerical study \cite{guimaraes2020direct} on turbulent planar jets of viscoelastic liquids has provided a deeper understanding on the experimental observations above. 
Direct numerical simulations using the FENE-P model~\cite{guimaraes2020direct} reveal that increasing the Weissenberg number in turbulent viscoelastic jets induces three main differences compared to Newtonian turbulence: (i) Depletion of small-scale vorticity throughout the flow due to the contribution of the polymer chains to vorticity transport, acting as a ``counter torque suppressing the generation of vortices'' \cite{guimaraes2020direct}, (ii) a tendency for eddy structures to become more elongated due to an increase in extensional viscosity, and (iii) a smaller rate of jet spreading due to the reduction in small-scale vorticity that lowers the viscous dissipation rate \cite{guimaraes2020direct, abreu2022turbulent}. Self-similar solutions were proposed for the rate of  jet spreading, the centerline velocity and the maximum polymer stresses using a similarity analysis that is independent of the specific viscoelastic model used for the polymer stress tensor \cite{guimaraes2020direct}.
% Part of the kinetic energy transferred through the classical nonlinear energy cascade is stored by elastic deformation of the polymers such that less is available for driving small scale vortical motion, leading to a decrease in the viscous dissipation rate \cite{abreu2022turbulent, guimaraes2020direct}. 
Through experiments, we have recently identified a universal scaling for the power-law decay of the turbulent kinetic energy spectrum for EIT with an exponent of $-3$ in viscoelastic axisymmetric jets \cite{yamani2021spectral}, which corroborates scaling arguments \cite{hinch1977mechanical, vonlanthen2013grid} suggesting that in EIT the flexible polymer chains sustain a persistent time\textendash averaged rate of strain and elastic stress through their continuous extension and relaxation. 

\noindent In this study, we focus on the onset of instability and transition to turbulence in submerged planar jets of dilute aqueous polymer solutions. We experimentally demonstrate the nonmonotonic effect of fluid elasticity on jet stability and characterize the unique spatio-temporal features associated with EIT in the fully-developed turbulent state far from the nozzle. We use laser Doppler velocimetry (LDV) and high-speed schlieren imaging to document the self-similarity of the time-averaged angle of jet spreading and the decay in centerline velocity of the jets with distance from the nozzle, and reveal that the scaling coefficients are independent of fluid elasticity but different from those of a Newtonian jet.  We show that EIT generates coherent structures that are larger and more elongated in the streamwise direction compared to Newtonian turbulence. Finally, analysis of the fluctuations in the LDV and schlieren signals measured at the centerline of the jet and in the fully turbulent region far from the nozzle are used to investigate the turbulent kinetic energy spectrum of EIT.

\section{Experimental Methods}\label{sec:ExpMethods}

\subsection{High-speed stereo-schlieren imaging}

\begin{figure*}[tbhp]
\centering
% \includegraphics[width=0.99\textwidth]{images/SchlierenSetupv4.png}
% \includesvg[width=0.99\textwidth]{ImagesFinal/SchlierenSetup.svg}
% \includegraphics[width=0.99\textwidth]{ImagesFinal/SchlierenSetup.pdf}
\includegraphics[width=0.99\textwidth]{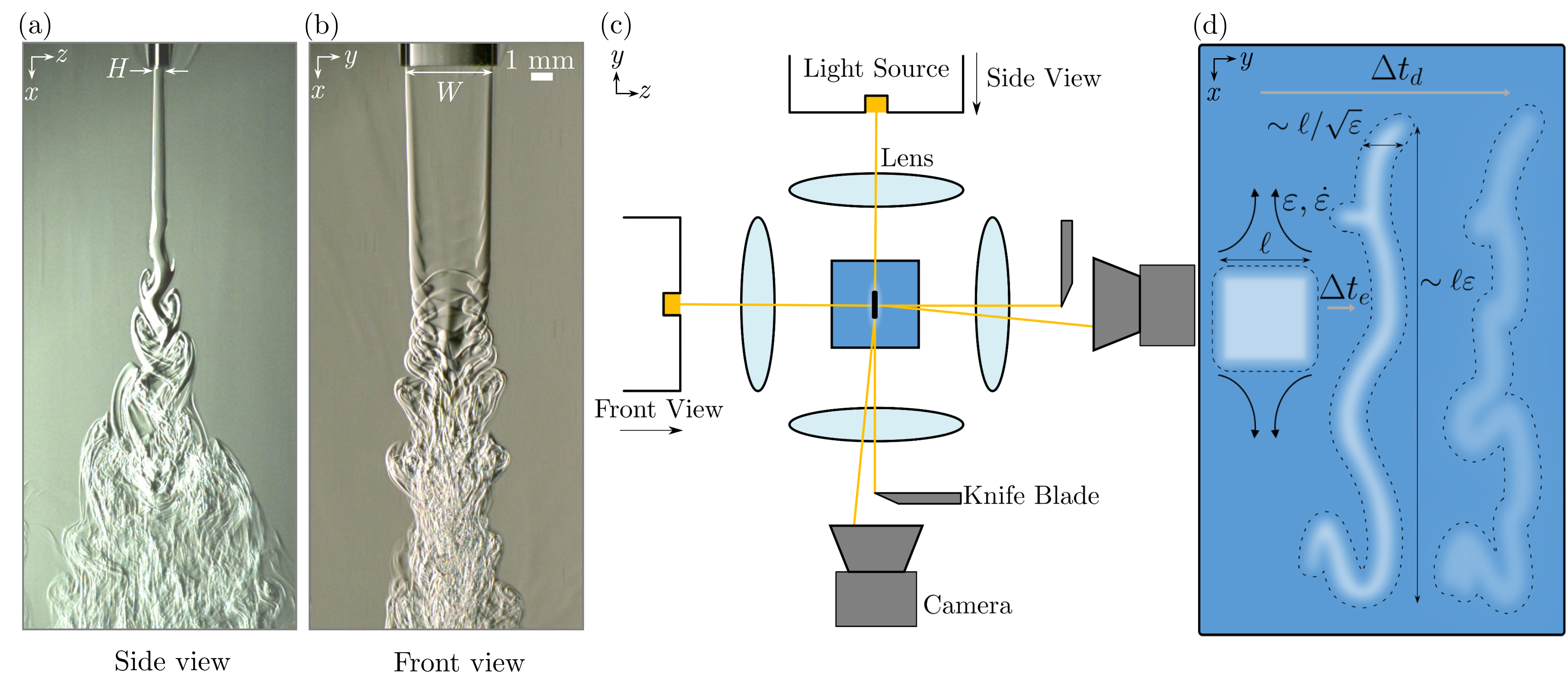}
\caption{(a) Side-view and (b) front-view of a turbulent Newtonian jet at $\textrm{Re} = 400$. \textcolor{black}{These views illustrate the jet for the streamwise range of $0\leq x/D_h \lesssim 50$}. (c) Top view schematic diagram of the high-speed stereo-schlieren imaging setup. (d)  An illustration of the straining of a material volume of dilute polymer solution with characteristic initial size $\ell$ (light blue square) in water (dark blue) as a result of a rapid strain $\varepsilon$ applied at a deformation rate $\dot{\varepsilon}$. The time scale for the extension is $\Delta t_e$ and the time scale for the diffusion of the polymer in the extended material volume due to $\varepsilon$ is $\Delta t_d$. If $\Delta t_e < \Delta t_d$, the evolution of the material volume can be visualized with schlieren imaging, provided that the undeformed material volume has a higher refractive index gradient than the minimum detectable refractive index gradient that sets the sensitivity of the schlieren setup.}
% \vspace{-14pt}
\label{fig:SchlierenSetup}
\end{figure*}

\noindent  A submerged planar jet of water or dilute polymer solution enters a large tank of quiescent water. A planar jet nozzle is generated from a precision glass capillary channel (Vitrocom-2540) with aspect ratio 1:10, where the short side of the capillary channel has an internal dimension $H = 0.4 \textrm{ mm}$ and the long side has an internal width of $W = 4 \textrm{ mm}$, as shown in Fig.\,\ref{fig:SchlierenSetup}(a) and (b). 
We define the hydraulic diameter of the nozzle as $D_h  = 2WH/(W+H) = 0.73 \textrm{ mm}$. For an ideal planar jet nozzle, in which $W\rightarrow \infty$, and with $H = 0.4 \textrm{ mm}$, $D_h \rightarrow 0.8 \textrm{ mm}$. The 1:10 aspect ratio capillary channel thus has a hydraulic diameter that is within $10\%$ of the hydraulic diameter of an ideal  two-dimensional planar channel. 
The nozzle exit is placed at depth $\sim 100D_h$ below the air-water interface of the tank. 

\noindent We use high-speed digital stereo-schlieren imaging to simultaneously visualize the front and side views of the planar jets, as schematically shown in Figure\,\ref{fig:SchlierenSetup}(c). We define the coordinate system such that the streamwise direction is in the $x$-direction and the spanwise directions in the front-view and side-view are in the $y$- and $z$-directions, respectively. We define the origin of the coordinate system at the center of the nozzle exit ($x = 0$) such that the exiting fluid stream spans the domain $-H/2 \leq z \leq H/2$ and $-W/2 \leq y \leq W/2$. Schlieren imaging reveals the boundaries of the fine-scale material regions that develop due to the refractive index gradient between the jet and the surrounding fluid as they are advected through the domain in a Lagrangian manner. In each of the two orthogonal schlieren lines shown in Fig.\,\ref{fig:SchlierenSetup}(c),
% (front-view visualization line shown in Fig.\,\ref{fig:SchlierenGain}a and side view visualization line shown in Fig.\,\ref{fig:SchlierenGainApp})
the lens closer to the light source collimates the light emitted from the extended LED light source. The collimated light rays are deflected from their original pathway if they pass through a medium with a locally different refractive index compared to the surrounding medium. The lens closer to the camera focuses the light rays on the edge of a knife blade that is adjusted vertically to block a portion of the deflected light rays, resulting in an image of the local light intensity gradient distinguishing the two fluid media (see Supplemental Material for Movies\,S1-6 \cite{supp}). 

\noindent To understand the capabilities and limitations of our visualization technique, it is helpful to establish criteria for successful schlieren imaging of the evolution of turbulent flow structures in dilute polymer solutions based on optical, fluid, and flow parameters. The sensitivity of a schlieren system is expressed in terms of the minimum detectable refractive index gradient~\cite{Settles2001}, 
\begin{equation}
    \left(\frac{\partial n}{\partial x_i}\right)_{min} = \left(\frac{\Delta E}{E}\right)\left(\frac{n_0}{L}\right)\left(\frac{a}{f_2}\right),
    \label{eq:sensitvity}
\end{equation}
where $x_i$ is the spanwise direction in each schlieren line ($y$ and $z$ for the front-view and side-view schlieren lines, respectively), $\Delta E/E$ is the minimum detectable light contrast (which is $\simeq 2\%$ for our setup), $n_0$ is the refractive index of the medium ($n_0 = n_{\textrm{water}} = 1.333$ in our experiments), $L$ is the length scale of the schlieren object being imaged in the direction of the optical axis, %($W = 4 \textrm{ mm}$ for the side view and $H = 0.4 \textrm{ mm}$ for the front-view), 
$a = 6 \textrm{ mm}$ is the length of the unblocked part of the light source (which is $10\%$ of the light source width in our experiments), and $f_2$ is the focal length of the second lens in each schlieren line ($f_2 = 0.3 \textrm{ m}$ and $f_2 = 0.15 \textrm{ m}$ for the front- and side-view schlieren lines, respectively). We calculate $\Delta E/E$ for our setup based on the standard deviation of each pixel due to experimental noise. In grayscale (8-bit) imaging, where each pixel has an intensity value between 0 to 255, calibration experiments show that the standard deviation of the pixel intensity when imaging the quiescent bath is $\pm 6$, thus $\Delta E/E = 6/256 \simeq 0.02$. This yields $(\partial n/\partial y)_{min} = 1.33 \textrm{ m}^{-1}$ and $(\partial n/\partial z)_{min} = 0.27 \textrm{ m}^{-1}$ for the front-view and side-view schlieren lines, respectively.    

\noindent To visualize a static structure with schlieren imaging, the refractive index gradient across the structure should be greater than the value $\left(\partial n / \partial x_i\right)_{min}$ for the schlieren setup. For a flow structure of characteristic size $\ell$, the refractive index gradient across the structure is denoted $\Delta n/\ell$. For a dilute polymer solution of concentration $c$, with $\partial n/\partial c$ a material property, we can write $\Delta n/\ell \sim (\partial n/\partial c)\times(c/\ell)$. The ratio of this value to the schlieren sensitivity denotes the static gain of the system. A static structure can thus be successfully visualized with schlieren imaging if the gain is 
\begin{equation}
\Gamma_{\textrm{static}}=\frac{\left(\partial n/\partial c\right)\left(c/\ell\right)}{\left(\partial n/\partial y\right)_{min}} > 1.
\label{eq:static_gain}
\end{equation}
The static gain can be calculated for the side-view and front-view visualizations using $H$ and $W$ as the characteristic lengths. A $1\%$ wt.\,water and sucrose solution is used for calibration and for Newtonian jet experiments due to the small density and viscosity difference with water ($<1\%$) and the known refractive index ($n_{\textrm{1\% wt.\,water-sucrose}} = 1.33445$) \cite{bin1988refractive}. For this solution, $\Delta n/H \simeq 3.63 \textrm{ m}^{-1}$ and $\Delta n/W \simeq 0.36 \textrm{ m}^{-1}$, yielding $\Gamma_{\textrm{static, side}}= 13.60 >1$ and $\Gamma_{\textrm{static, front}}= 0.27 < 1$. This is consistent with our visualization shown in Fig.\,\ref{fig:SchlierenSetup}(a) and (b). The fluid stream at the nozzle exit can be resolved from the background in the side-view visualization. However, only the lateral edges of the fluid stream at the nozzle exit  ($y \simeq \pm W/2$) are resolved from the background in the front-view visualization and the center of the fluid stream ($y \simeq 0$) has the same light intensity as the background fluid. The static gain criterion determines the size of the largest fluid structures that can be visualized for a particular polymer solution. However, it does not account for the effects of molecular diffusion and cannot predict the size of the finest flow structures that can be visualized with schlieren imaging while they undergo deformations in the flow.

\noindent To evaluate the ability to visualize the dynamic evolution of flow structures with schlieren imaging, we consider an arbitrary material volume element of characteristic size $\ell$ (Fig.\,\ref{fig:SchlierenSetup}(d)). A strain rate $\dot{\varepsilon}$ imposed on this material volume element for a time $\Delta t_e$ leads to a material stretching with true strain $\varepsilon = \dot{\varepsilon}\Delta t_e$. In a uniaxial extension of the material volume ($x$-direction in Fig.\,\ref{fig:SchlierenSetup}(d)) and assuming $\varepsilon \gg 1$, the material volume is extended from an initial length $\ell$ to a final length $\sim \varepsilon \ell$. Conservation of mass \textcolor{black}{together with incompressibility} imply a decrease in the lateral size of this material volume element from $\ell$ to $\sim \ell/\sqrt{\varepsilon}$ in the orthogonal directions ($y$- and $z$-directions in Fig.\,\ref{fig:SchlierenSetup}(d)). We seek to understand how this straining affects the distribution and visibility of the polymer solution in the flow. The derivative of the refractive index with respect to the polymer concentration $\partial n/ \partial c = \left(\partial n/ \partial y\right) \big/ \left(\partial c/ \partial y\right)$ is a material property and constant for a specific polymer solution. It has to remain constant at all times; an increase in $\partial n/ \partial y$ results in an increase in $\partial c/ \partial y$. An increase in the refractive index gradient, $\partial n/ \partial y$, is favorable for visualization as it increases the ratio $ (\partial n/ \partial y) \big/ (\partial n /\partial y)_{min}$. An increase in the concentration gradient,  $\partial c/ \partial y$, however, also increases the diffusive flux of the polymer into the background field, which leads to a spreading of the material volume and a progressive homogenization of the polymer concentration. We use Fick's second law to quantify this enhanced diffusive flux, 
\begin{equation}
    \frac{\partial c}{\partial t} = D \frac{\partial^2 c}{\partial y^2}, 
    \label{eq:fick}
\end{equation}
where $D$ is the diffusion coefficient of a polymer in a solvent. From Eq. \ref{eq:fick}, we expect a characteristic diffusion time scale $\Delta t_d \sim \Delta y^2/D$. From Zimm theory \cite{rubinstein2003polymer} the diffusion coefficient is given by 
\begin{equation}
    D = 0.196\frac{k_BT}{\eta_s a N^{\nu}}, 
\end{equation}
where $k_B$ is the Boltzmann constant, $T$ is the temperature, $\eta_s$ is the solvent viscosity, $a$ is the length of the polymer chain repeat unit ($a=0.28$ nm for polyethylene oxide (PEO)), $N$ is the number of repeat units in the polymer chain, and $\nu$ is the solvent quality parameter ($\nu \simeq 0.6$ for PEO in water). The number of repeat units in a polymer chain scales linearly with the polymer molecular weight ($M_w = 44.05N+18.02$ for PEO \cite{brandrup1989polymer}). We thus expect $D \sim N^{-\nu} \sim M_w^{-\nu}$. Using the scaling for $D$ and knowing that $\Delta y \sim \ell/\sqrt{\varepsilon}$, the scaling for the polymer diffusion time scale can be expressed as $\Delta t_d \sim \ell^2/\varepsilon D \sim M_w^{\nu}$.

\noindent The dynamic evolution of a material volume in the flow can be visualized with schlieren imaging only if the enhanced diffusive flux resulting from the steepened concentration gradients in the rapidly stretched material volume is slower than the rate of accumulation in strain. This ratio of time scales is expressed by the P\'eclet number,
\begin{equation}
    \textrm{Pe} = \frac{\Delta t_d}{\Delta t_e} \sim \frac{(\ell/\sqrt{\varepsilon})^2/D}{\varepsilon / \dot{\varepsilon}} = \frac{\dot{\varepsilon} \ell^2}{\varepsilon^2 D},
    \label{eq:peclet}
\end{equation}
which compares the straining time scale of the arbitrary material volume to the enhanced diffusion time scale arising from the large polymer concentration gradients that result from the  straining. For $\textrm{Pe}>\textrm{O(1)}$, the strain of the material element from turbulent velocity fluctuations of strength $\dot{\varepsilon}$ occurs faster than the molecular diffusion. The polymer concentration gradients in the flow thus persist, enabling the evolving Lagrangian material volume to be visualized with schlieren imaging, provided that the imaging speed of the camera is fast enough. This additional criterion expresses a dynamic gain for the schlieren system and acts as a condition for visualizing flow structures of characteristic size $\ell$ in a rapidly fluctuating flow as they undergo straining deformations of magnitude $\varepsilon$ due to strain rates of magnitude $\dot{\varepsilon}$. The dynamic gain can be written as \begin{equation}
     \Gamma_{\textrm{dynamic}} = \left(\frac{\dot{\varepsilon} \ell^2}{\varepsilon^2 D}\right) > 1.
     \label{eq:dynamic_gain}
\end{equation}

\noindent \textcolor{black}{We note that the dynamic gain expressed in Eq.\,(\ref{eq:dynamic_gain}) is only a scaling argument for visualization of unstable and turbulent jets since (i) Fick's second law is strictly only valid under no-flow condition and (ii) a jetting flow is not homogeneous uniaxial extension.} A material volume element of initial size $\ell \sim D_h = 0.73 \textrm{ mm}$ that is extended in the $x$-direction by $\Delta \ell/\ell = \varepsilon \sim 10$ at a Reynolds number of $\textrm{Re} = 100$ experiences a strain rate of $\dot{\varepsilon} \sim U_0/D_h \sim 100  ~ 1/\textrm{s}$, where $U_0$ is the mean flow velocity at the nozzle.  Considering  $D \simeq 10^{-9} \textrm{ m}^2\textrm{/s}$ for a $1\%$ wt.\,water-sucrose solution \cite{price2016sucrose}, the dynamic gain is $\Gamma_{\textrm{dynamic}} = \textrm{Pe} = \Delta t_d/\Delta t_e \sim 1000$. \textcolor{black}{Given that our scaling for the dynamic gain is at least three orders of magnitude higher than unity}, visualization of the evolution of material elements undergoing strong deformations in the flow is possible without the need for additional optical magnification. \textcolor{black}{If the dynamic gain is set to unity, the smallest length scale that can be visualized for these flow conditions, \textit{i.e.} the theoretical resolution, is $2.3\times 10^{-5} ~ \textrm{m}$, which is close to our optical resolution, $2.7\times 10^{-5} ~ \textrm{m}$.} The diffusion coefficient of the polymers used in this work is three orders of magnitude smaller than that of water ($D \sim 10^{-12} \textrm{ m}^2\textrm{/s}$), which further enhances the dynamic gain. These dynamic gain calculations are consistent with Fig.\,1(a) and (b), in which we are able to visualize fine-scale Lagrangian structures in the turbulent flow despite their enhanced diffusion rate.

\subsection{Laser Doppler velocimetry}

\noindent Schlieren imaging visualizes the evolution of spatio-temporal flow structures in a Lagrangian manner without the need for tracing particles. It is, however, difficult to extract local velocity profiles of the flow and their fluctuations from schlieren images. We use laser Doppler velocimetry (LDV), an Eulerian method that measures the velocity and velocity fluctuations with high temporal resolution at a fixed Eulerian location in the flow.  

\noindent We use the miniLDV G5B sensor from MSE Inc.. Two laser beams collide at the stand-off distance of the sensor, creating an ellipsoidal probe volume containing multiple interference fringes. The probe volume has a length of 1.5~mm and a diameter of $200~\mu\textrm{m}$ at full width half maximum of the laser intensity profile.
% , as shown schematically in Fig.\,\ref{Miro S302, Phantomfig:LDVApp}
The jet is seeded with 10~ppm of Titanium(IV) oxide powder (Sigma-Aldrich, 224227) with grain size smaller than $5~\mu\textrm{m}$. The particles scatter light as they pass through the bright fringes in the ellipsoidal probe volume. This scattering generates a sinusoidal intensity burst that is detected by a photodetector. \textcolor{black}{The frequency of the Doppler-shifted scattered signal, $f_s(x,z,t)$, at Eulerian location $(x,z)$ and time $t$, is calculated in real time by applying a bandpass filter and using a fast Fourier transform (FFT) algorithm. Knowing the spacing between fringes, $s = 8.55~\mu \textrm{m}$, the instantaneous velocity of the particle is calculated as
\begin{equation}
    \tilde{U} = f_s(x,z,t) \times s. 
\end{equation}
The velocity is decomposed into a mean component, $U(x,z)$, and a fluctuating component, $u(x,z,t)$, with zero mean: 
\begin{equation}
    \tilde{U}(x,y,t) = U(x,z) + u(x,z,t).
\end{equation}
We align the two beams in the vertical $xy$-plane at the centerline of the jet ($z=0$) to measure the streamwise velocity at the centerline of the jet: $\tilde{U}_{cl}(x,t) = U_{cl}(x)+u_{cl}(x,t)$.}   

\begin{figure*}[tbhp]
\centering
\includegraphics[width=0.99\textwidth]{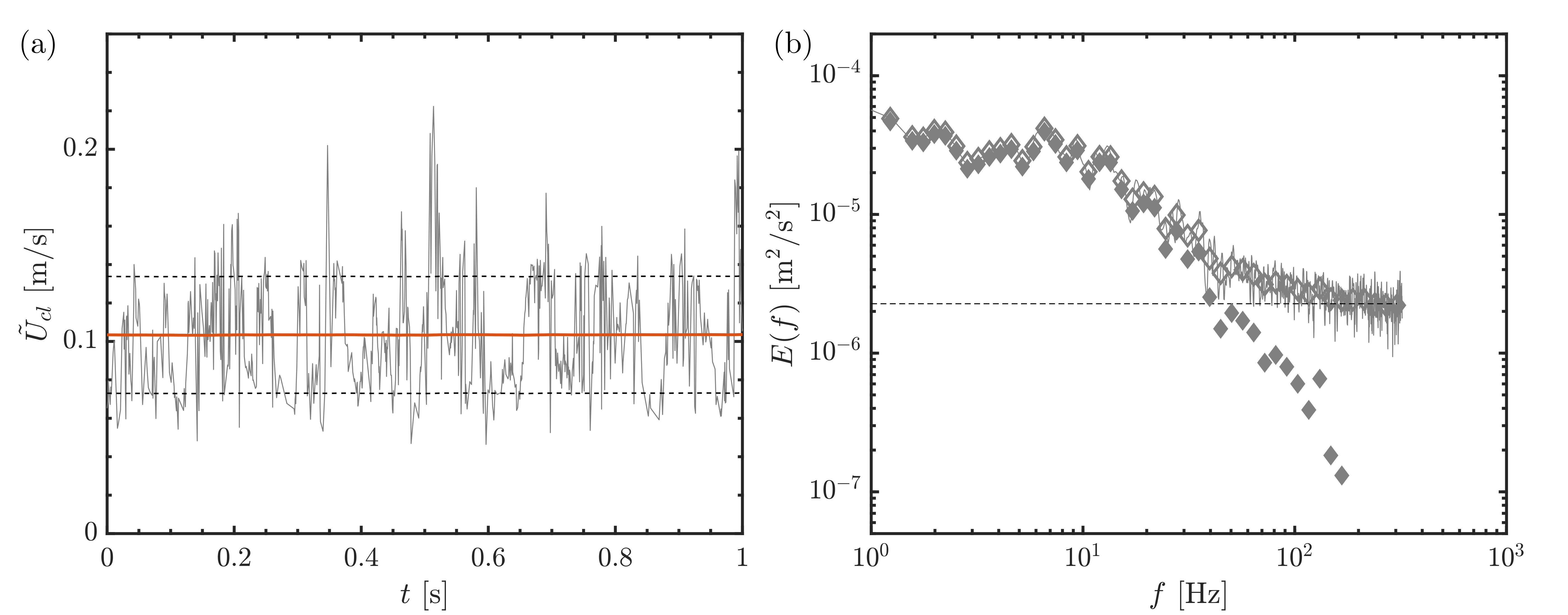}
\caption{
% (a) Schematic of our LDV setup. The intersection of two laser beams generates fringes in a 3D ellipsoidal probe volume with a length of 1.5 mm and a diameter of 200 $\mu$m at full width half maximum.
(a) A section of a velocity time series, $\tilde{U}_{cl}(t)$, at the streamwise location $x = 40D_h$ from the nozzle and at the centerline of the jet for a Newtonian turbulent jet at $\textrm{Re} = 400$. The average data rate is 700 bursts/second. The red solid line denotes the mean centerline velocity, $U_{cl}$, at $x = 40D_h$, the black dashed lines denote one standard deviation above and below the mean velocity, $U_{cl} \pm u_{cl}^{RMS}$, computed from the entire record of $10,000$ bursts collected over 16 seconds. (b) PSD of the velocity fluctuations. The thin solid line denotes the PSD calculated from the LDV velocity measurements. The hollow diamonds ($\MyDiamond[draw=water,fill=white]$) denote the PSD averaged over 55 logarithmically spaced bins ($\sim 22$ bins per decade), the dashed line indicates the noise floor of the LDV measurement given by a weak power-law $E_{\textrm{noise}} \sim f^{-0.04}$, and the filled diamonds ($\MyDiamond[draw=water,fill=water]$) denote the noise-corrected power spectrum averaged over logarithmically spaced bins.}
% \vspace{-14pt}
\label{fig:LDV}
\end{figure*}

\noindent The PSD of the velocity fluctuations, also referred to as the turbulent kinetic energy spectrum $E(f)$, provides information about the nature of turbulence. A typical section of a velocity time series is shown in Fig.\,\ref{fig:LDV}(a) for a Newtonian turbulent jet at $\textrm{Re} = 400$, from which we calculate the PSD shown as a solid line in Fig.\,\ref{fig:LDV}(b). Additional details on the calculation of the PSD from signals of finite duration are provided in Appendix\,\ref{app:psd}.

\noindent The measured PSD approaches a plateau at high frequencies that indicates the noise floor inherent to the measurement. The sources of noise include velocity variations within the LDV probe volume, unwanted reflections of the laser beams from surfaces, and the limited sensitivity of the photodetector. The noise level can be subtracted to remove its contribution to the power spectrum, yielding the turbulent kinetic energy power spectrum \cite{warholic1997phdthesis, warholic1999influence, niederschulte1990measurements, gunther1998turbulent, adrian1986power}. 
%It should be noted that in most cases, the noise plateau is the result of white Gaussian noise and hence shows itself as a plateau in the power spectrum. It is, however, possible to have a non flat, \textit{i.e.}, tilted or smoothly varying noise spectrum. 
We follow the approach suggested in Chapter 13 of \cite{press1992numerical} and fit a power law to the high frequency, noise-dominated region of the spectrum ($f \geq 100 \textrm{ Hz}$). This fit can be extrapolated to lower frequencies where the turbulent energy signal is dominant, and is then subtracted across the entire spectrum to access the noise-corrected turbulent kinetic energy power spectrum, as shown by the filled diamonds in Fig.\,\ref{fig:LDV}(b). 

%\noindent The power-law fit approach is implemented to the power spectrum show in Fig.\,\ref{fig:LDV}b with solid line and hollow symbols. The nature of noise in our system is mainly white Gaussian and the power-law fit has a slope that is very close to zero resulting in a flat noise floor shown with dashed line in Fig.\,\ref{fig:LDV}b. The subtraction of the noise floor from the power spectrum reveals the turbulent kinetic energy power spectrum shown with filled symbols in Fig.\,\ref{fig:LDV}b.

\subsection{Dimensionless numbers and rheological characterization}

\noindent The dilute polymer solutions studied in this work are formulated from polyethylene oxide (PEO) (Sigma-Aldrich) dissolved in deionized water (Sigma-Aldrich). Five solutions are prepared using PEO with molecular weights $M_w$=$4\times 10^6 \textrm{ g/mol}$ and $M_w$=$8\times 10^6 \textrm{ g/mol}$. The PEO concentration $c$ is kept at $c/c^*<1$, where $c^*$ is the polymer overlap concentration, to ensure minimal chain-chain interactions. Stock solutions of $0.5\%$ wt.\,of each molecular weight are gently mixed for at least 10 hours using a roll-mixer to avoid polymer degradation. 
%on a Wheaton Roller Culture Apparatus (RLR 348922)
%using roll-mixers for 48 to 240 hours depending on the molecular weight %(higher rolling time is needed for polymers with higher molecular weights)
%until a uniform viscoelastic fluid is achieved. 
Samples at different concentrations are prepared by dilution with deionized water. To attain the required refractive index gradient for schlieren
imaging, 1\% wt.\,sucrose is added to all solutions. The addition of sucrose changes the viscosity by $\sim 1$\% and does not affect the elasticity of the solutions.
% The details of each sample and the symbols used to refer to them in the figures of this work are shown in Table \ref{tab:propertySI}. 
% Uniform dispersion can be assessed through optical transmission and refractive index measurements. 
% Experiments were carried out in less than 24 hours after sample preparation to avoid polymer degradation.

\noindent The dimensionless numbers characterizing the jet are (i) the Reynolds number $\textrm{Re} = \rho U_0 D_h/\eta_0$, where $\rho$ is the fluid density, $U_0$ the mean flow velocity at the nozzle, $D_h$ the hydraulic diameter of the nozzle, and $\eta_0$ the zero shear viscosity of the solution, (ii) the elasticity number $\textrm{El} = \eta_p\lambda/\rho (D_h/2)^2$, where $\eta_p = \eta_0-\eta_s$ is the polymer viscosity and $\lambda$ the extensional relaxation time, and (iii) the polymer chain extensibility $L_{max} = r_{max}/\langle r_0^2\rangle ^{1/2}$, 
where $r_{max} \sim M_w$ is the length of a fully extended polymer chain, $\langle r_0^2 \rangle ^{1/2} \sim M_w^{\nu}$ is the equilibrium root-mean-square end-to-end separation of the polymer in the coil state and $\nu$ is the solvent quality parameter \cite{rubinstein2003polymer}. From these dimensionless numbers and following previous work on planar jets \cite{guimaraes2020direct}, a Weissenberg number can be defined as $\textrm{Wi} = \lambda U_0 /H \simeq \lambda U_0 / (D_h/2)$. Assuming $D_h \simeq 2H$ due to the large aspect ratio of the nozzle, this yields $\textrm{El} = 2\left(1-\beta \right) \left(\textrm{Wi}/\textrm{Re}\right)$, where $\beta = \eta_s/\eta_0$ is the solvent viscosity ratio. 
% (where $R_{max} \sim M_w$ is the length of a fully extended polymer chain, $\langle r_0^2 \rangle ^{1/2} = \sqrt{6} R_g \sim M_w^{\nu}$ is the equilibrium root-mean-square end-to-end separation of the polymer in the coil state, $R_g$ is the equilibrium radius of gyration of polymer molecule in the coil state, and $\nu$ is the solvent quality parameter \cite{rubinstein2003polymer}).

\noindent %To systematically vary the viscoelasticity, we use five dilute polymer solutions with different polymer molecular weights ($M_w$) and concentrations ($c$). 
%The rheological properties of the solutions are summarized in Table\,\ref{tab:propertySI}. %The zero shear viscosity and the extensional relaxation time are measured experimentally (boldfaced values in Table\,\ref{tab:propertySI}). Here we show the experimental protocols for the rheological measurements and our approach for calculating the material properties that not determined experimentally. 

\noindent The rheological properties of the solutions are summarized in Table\,\ref{tab:propertySI}. The zero shear viscosity $\eta_0$ is measured using an Ubbel\"{o}hde-type suspended-level capillary viscometer (size 0B, Cannon Instrument) immersed in an isothermal water bath at $25^{\circ}$C. %We adopt the experimental procedure from \cite{standard2006d445}, where based on their calculation of the tolerance band the calculated error for a zero shear rate dynamic viscosity measurement is $ \pm 0.01 \textrm{ mPa.s}$. 
We calculate the reduced viscosity $\eta_{red}$ as \cite{pamies2008determination, Rajappan_2019} 
\begin{equation}
    \eta_{red}(c) = \frac{(\eta_0-\eta_s)}{\eta_s c},
\end{equation}
where $\eta_s$ is the solvent viscosity and $c$ is the polymer concentration. From the Huggins relation \cite{pamies2008determination, Rajappan_2019} we expect
\begin{equation}
    \eta_{red} = [\eta] + k_H [\eta]^2 c + O(c^2),
\end{equation}
% $\eta_{red} = [\eta] + k_H [\eta]^2 c + O(c^2)$, 
where $[\eta]$ is the intrinsic viscosity and $k_H$ the Huggins constant. %This linear approximation is accurate for our dilute solutions \cite{pamies2008determination, Rajappan_2019}. 
We extrapolate this linear regression to zero concentration to find the intrinsic viscosity of the polymer in dilute solution. 
% Figure\,\ref{fig:IntrinsicViscosity}a shows the intrinsic viscosity of PEO samples studied in the present work (red filled circles) and compares them to the values previously reported in the literature \cite{christanti2002effect, ferguson1992break, stokes1998swirling, Tirtaatmadja2006} (black hollow symbols). The 90\% confidence bounds of the linear regression determine the error bar for each data point.
The uncertainty in the intrinsic viscosity increases as the molecular weight of the polymer increases, primarily due to the increase in polydispersity. 
% The solid line in Fig.\,\ref{fig:IntrinsicViscosity}a denotes the previously reported form of the Mark–Houwink–Sakurada equation \cite{bird1987dynamicsv1} for PEO in water.

\begin{table}[thbp]
\centering
\caption{\label{tab:propertySI}Composition and material parameters of the polymer solutions at 25$^{\circ}$C.}
\resizebox{0.99\textwidth}{!}{
\begin{threeparttable}
\begin{tabular}{cccccccccc} 
\hline\hline
\\[-1em]
 $M_w \textrm{ [g/mol]}$          & $~~L_{max}~~$             & $~~\left[\eta\right] \textrm{ [dl/g]}~~$                                                     & $~~c^* \textrm{ [ppm]}~~$  & $~~c \textrm{ [ppm]}~~$ & $~~c/c^*~~$ & $~~\eta_0 \textrm{ [mPa.s]}~~$                                                       & $~~\lambda \textrm{ [ms]}~~$ & $~~\textrm{El}~~$                                       & $~~\beta~~$  \\ 
 \\[-1em]
\hline
\\[-1em]
\multirow{4}{*}{$4\times 10^6$ } & \multirow{4}{*}{$68.4$} & \multirow{4}{*}{$\mathbf{12.1\pm 0.8}$\tnote{$\textrm{a}$}} & \multirow{4}{*}{$547$\tnote{$\textrm{b}$}} & $~50~~$\,\mysquare{4M50ppm}                  & $0.09$    & $\mathbf{0.98\pm 0.01}$\tnote{$\textrm{c}$}                                                                  & $\mathbf{~5 \pm 1}\tnote{$\textrm{d}$}$      & $\left(4.0 \pm 1.0 \right)\times 10^{-3}$   & $0.97$                                  \\
                                 &                       &                                                                                          &                         & $100~~$\mysquare{4M100ppm}                  & $0.18$    & $\mathbf{1.04}$                                                                  & $\mathbf{13\pm 3}$      & $\left(1.3 \pm 0.3 \right)\times 10^{-2}$   & $0.88$                                   \\ 
                                 &                       &                                                                                          &                         & $200~~$\mysquare{4M200ppm}                  & $0.37$    & $\mathbf{1.19}$                                                                  & $\mathbf{21\pm 1}$      & $\left(4.1 \pm 0.3 \right)\times 10^{-2}$   & $0.78$                                   \\ 
                                 &                       &                                                                                          &                         & $300~~$\mysquare{4M300ppm}                  & $0.55$    & $\mathbf{1.35}$                                                                  & $\mathbf{~~\,23\pm0.5}$      & $\left(6.6 \pm 0.2 \right)\times 10^{-2}$   & $0.70$                                   \\ 
\hline
\multirow{2}{*}{$8\times 10^6$}  & \multirow{2}{*}{$91.4$} & \multirow{2}{*}{$\mathbf{15\pm 5}$\tnote{$\textrm{a}$}}      & \multirow{2}{*}{$348$} &\multirow{2}{*}{$150~$ \mycircle{8M150ppm}}                & \multirow{2}{*}{$0.43$}    & \multirow{2}{*}{$\mathbf{1.28}$}                                                                  & \multirow{2}{*}{$\mathbf{54\pm 1}$}       & \multirow{2}{*}{$\left(1.24 \pm 0.4 \right)\times 10^{-1}$}   & \multirow{2}{*}{$0.75$}                                  \\
\\
\hline\hline
\end{tabular}
\begin{tablenotes}\footnotesize
\item[*] Bold faced values are determined experimentally.
% \item[a] Calculated using the Mark–Houwink–Sakurada equation for PEO, $[\eta] = 0.00072 \times M_w^{0.65}$ \cite{Tirtaatmadja2006}.
% \item[b] Calculated based on zero shear viscosity $(\eta_0)$ measurements from capillary viscometry.
\item[a] Calculated based on zero shear viscosity $(\eta_0)$ measurements from capillary viscometry. Intrinsic viscosity can also be calculated using the Mark–Houwink–Sakurada equation for PEO, $[\eta] = 0.00072 \times M_w^{0.65}$ \cite{Tirtaatmadja2006}, which provides 14.08 dl/g and 22.1 dl/g for molecular weights $4\times 10^6$ g/mol and $8\times 10^6$ g/mol, respectively.
\item[b] Calculated using $c^* = 0.77/[\eta]$, where $[\eta]$ is the intrinsic viscosity \cite{Tirtaatmadja2006}. 
\item[c] Measured using Ubbel\"{o}hde-type suspended-level capillary viscometer (size 0B, Cannon Instrument) immersed in isothermal water bath at $25^{\circ}$C.
% \item[e] Calculated using the best fit to relaxation times measured using CaBER, $\lambda/\lambda_{Zimm} = (30\pm4)\left(c/c^*\right)^{0.6\pm0.1}$. 
\item[d] Measured using CaBER. 
\end{tablenotes}
\end{threeparttable}
}

\end{table}

\begin{figure*}[tbhp]
\centering
\includegraphics[width=0.99\textwidth]{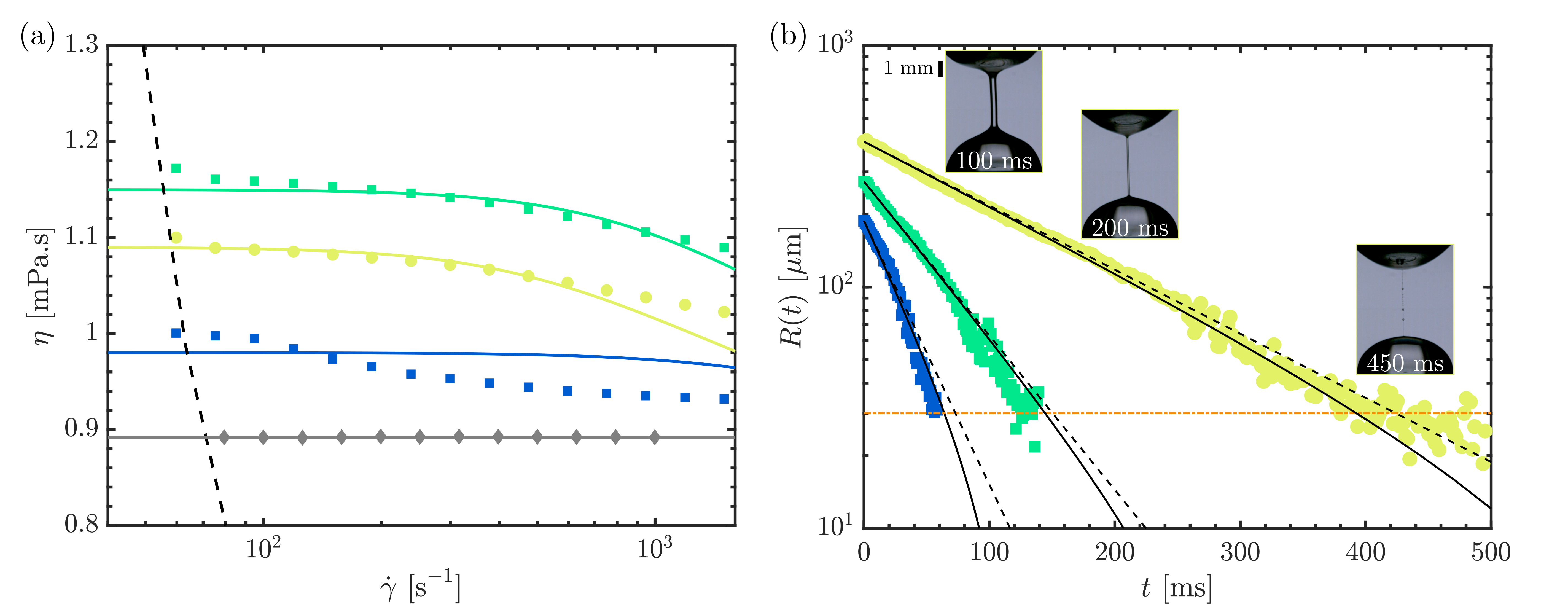}
\caption{
% (a)Intrinsic viscosities of polyethylene oxide (PEO) solutions used in this work calculated based on capillary viscometry measurements (filled symbols) and intrinsic viscosities reported in the literature \cite{christanti2002effect, ferguson1992break, stokes1998swirling, Tirtaatmadja2006} (hollow symbols). The solid line is the Mark–Houwink–Sakurada equation for PEO solutions, $[\eta] = 0.00072 \times M_w^{0.65}$ \cite{Tirtaatmadja2006}, the dashed line denotes an expression for the Mark–Houwink–Sakurada equation for PEO for low/moderate molecular weights $M_W \leq 3 \times 10^4 \textrm{ g/mol}$ \cite{brandrup1989polymer}. 
(a) Shear viscosity of deionized water ($\MyDiamond[draw=water,fill=water]$) and the highest concentration PEO solutions used in this work (for which maximum shear thinning is expected), \textit{i.e.}, $\textrm{El} = 0.013$ (\mysquare{4M100ppm}), $\textrm{El} = 0.066$ (\mysquare{4M300ppm}), and $\textrm{El} = 0.0124$ (\mycircle{8M150ppm}).%, and comparison with measured  shear viscosity of deionized water. 
The solid lines denote the shear viscosity predicted by the FENE-P model for the parameters reported in Table\,\ref{tab:propertySI}. The dashed line denotes the low torque limit of the instrument corresponding to $\tau_{min} = 0.1 \textrm{ }\mu\textrm{N.m}$. (b) Evolution of the filament radius at the thinnest axial location, $R(t)$, during filament thinning experiment for the same PEO solutions as in (a). The symbols represent experimental data obtained by image processing of high-speed videos taken in a capillary breakup extensional rheometry (CaBER). Dashed and solid lines correspond to predictions of the Oldroyd-B and FENE-P \cite{bird1987dynamicsv1} models, respectively. The dash-dot line denotes the filament radius at which beads-on-a-string structures \cite{oliveira2006iterated} start to appear. The insets show snapshots of the filament profiles for a PEO solution with $M_w = 8\times 10^6 \textrm{ g/mol}$ ($L_{max} = 91.4$) and $c = 300$ ppm.}
\label{fig:Rheology}
\end{figure*}

\noindent Rheological measurements with these dilute polymer solutions indicate that they are weakly shear thinning, as shown in Fig.\,\ref{fig:Rheology}(a), where we report viscosity data for the range of shear rates that the jets experience in the schlieren experiments ($\dot{\gamma} \sim U_0/D_h \leq 1000 \textrm{ s}^{-1}$). To increase the magnitude of the torque signal, we use a double-wall concentric cylinder geometry on a controlled stress rheometer (DHR3, TA Instruments). %The dashed line denotes the minimum torque limit of the rheometer (corresponding to $\tau_{min} = 0.1 \textrm{ }\mu\textrm{N.m}$). 
Given the weakness of the shear thinning (less than 10\% for all samples) we here neglect these effects, following previous experimental studies \cite{Samanta2013, Choueiri2018} and report material quantities and dimensionless parameters in terms of zero shear-rate properties. The shear viscosity of our solution was also calculated using the finitely extensible nonlinear elastic-Peterlin (FENE-P) \cite{bird1987dynamicsv1} model with the parameters reported in the legend of Fig.\,\ref{fig:Rheology}(a). The predictions are in good agreement with the experimental results for the two solutions with higher polymer concentrations (note the linear scale of the ordinate). The FENE-P model, however, fails to accurately predict the slow and protracted weak shear thinning behavior of the lowest concentration solution. This mismatch potentially arises due to the previously-reported weak aggregation of PEO molecules in ultra-dilute polymer solutions that is slowly disrupted by imposed shear, which cannot be modelled by a simple non-interacting dilute polymer solution model such as the FENE-P dumbbell model \cite{shetty2009aggregation}. Regardless of this small (less than 1\%) quantitative mismatch between experimental and FENE-P model results at the lowest polymer concentration, the small extent of shear thinning predicted by the FENE-P model supports our decision to neglect shear-thinning effects.

% In agreement with our experimental results, calculation of shear viscosity of our solutions using the finitely extensible nonlinear elastic-Peterlin (FENE-P) \cite{bird1987dynamicsv1} model and with the parameters shown in the legend of Fig.\,\ref{fig:Rheology}a, further confirms the negligible shear thinning of the studied solutions.    

\noindent The extensional relaxation times of the dilute PEO solutions are measured using a capillary breakup extensional rheometer (CaBER) \cite{oliveira2006iterated}. As the two plates of the CaBER device are separated by a step function, a filament of the viscoelastic test liquid is formed between the plates, which becomes progressively thinner with time under the action of capillary pressure (Fig.\,\ref{fig:Rheology}(b)). A high-speed video is recorded and the evolution of the midpoint radius of the viscoelastic fluid filament with time is measured by an edge-detection algorithm. The extensional relaxation time of the polymer solution is calculated from an elasto-capillary balance on the thinning fluid filament, which becomes appropriate on small length scales~\cite{oliveira2006iterated}.
% for the range of evolution where elastic and capillary forces are the dominant forces. 
The evolution of the filament radius with time in the elasto-capillary region is given by 
\begin{equation}
    \frac{R(t)}{R_0} = \left(\frac{\eta_p R_0}{2\lambda \sigma}\right)^{1/3} \exp \left(-\frac{t}{3\lambda}\right),
    \label{eq:Oldroyd-B}
\end{equation}
% $R(t)/R_0  = (\eta_p R_0 /\lambda \sigma)^{1/3} \exp (-t/3\lambda)$, 
where $R(t)$ is the midpoint radius of the elastic filament, $R_0$ is the initial radius of the filament at the start of the elasto-capillary regime, $\eta_p = \eta_0 - \eta_s$ is the polymer contribution to the total viscosity, $\sigma$ is the surface tension of the polymer solution, and $\lambda$ is the characteristic extensional relaxation time of the polymer solution \cite{clasen2006beads}.

\noindent To illustrate how the CaBER results compare with the predictions of the Oldroyd-B and FENE-P models \cite{bird1987dynamicsv1}, the evolution of the elastic filament radius over time is plotted in Fig.\,\ref{fig:Rheology}(b). We follow the derivation in \cite{wagner2015analytic} for the evolution equation of the filament radius with time based on the Oldroyd-B and FENE-P models in the elastocapillary regime (see Appendix \ref{app:FENE-P-CaBER} for details). The critical radius below which the ligament is no longer a uniform elastic thread but develops a beads\textendash on\textendash a\textendash string morphology \cite{clasen2006beads} is shown with a horizontal dash-dot line. \textcolor{black}{It should be noted that while the kinematics of a planar jet are shear dominated, it is far from steady simple shear or uniaxial extension that are used here to characterize the rheological properties of the solutions. Rheological characterization under conditions of homogeneous prescribed deformation, however, is indispensable for reproducing our jet experiments or for performing complementary numerical simulations and data assimilation \cite{zaki_limited_2021,wang_jfm2019,du_VarPINN2022}.}%This radius is in agreement with the values reported in \cite{oliveira2006iterated}.

\noindent To calculate the polymer chain extensibility $L_{max} = r_{max}/\langle r_0^2\rangle ^{1/2}$, the length of a fully extended polymer chain, $r_{max}$, is estimated using $r_{max} = lM_w/m_0$, where $l~=~0.28 \textrm{ nm} $ is the length and $m_0 = 44 \textrm{ g/mol}$ the molar mass of the ethylene oxide repeat unit \cite{burshtein2017inertioelastic}. The equilibrium root-mean-square end-to-end separation of the polymer in the coil state is $\langle r_0^2 \rangle ^{1/2} = \sqrt{6} R_g \sim M_w^{\nu}$, where $\nu$ is the solvent quality parameter, and $R_g$ is the equilibrium radius of gyration of the polymer in the coil state \cite{devanand1991asymptotic}.

\section{Results and Discussion}\label{sec:Results}

\subsection{Stability of Newtonian and viscoelastic planar jets}\label{sec:DMD}

\noindent As the Reynolds number increases, our high-speed schlieren images show that the unperturbed jets transition from a laminar state to an unsteady state and ultimately to a fully turbulent state, as shown in Supplemental Material, Movies S1-S6 \cite{supp} and Appendix\,\ref{app:snapshots}. At a Reynolds number $\textrm{Re} = 100$, the jets show the evolution and growth of disturbances but have not yet transitioned to a turbulent state over the field of view ($x \leq 50D_h$). An increase in the elasticity number first makes the planar jet more unstable, but then subsequently partially stabilizes the jet at higher elasticity numbers, as seen in the front-view schlieren snapshots in  Fig.\,\ref{fig:DMDSnapshots} and in the Supplemental Material, Movie\,S7 \cite{supp}. To quantify these observations we use dynamic mode decomposition (DMD) \cite{schmid2010dynamic}.

\begin{figure*}[htbp]
\centering
\includegraphics[width=0.99\textwidth]{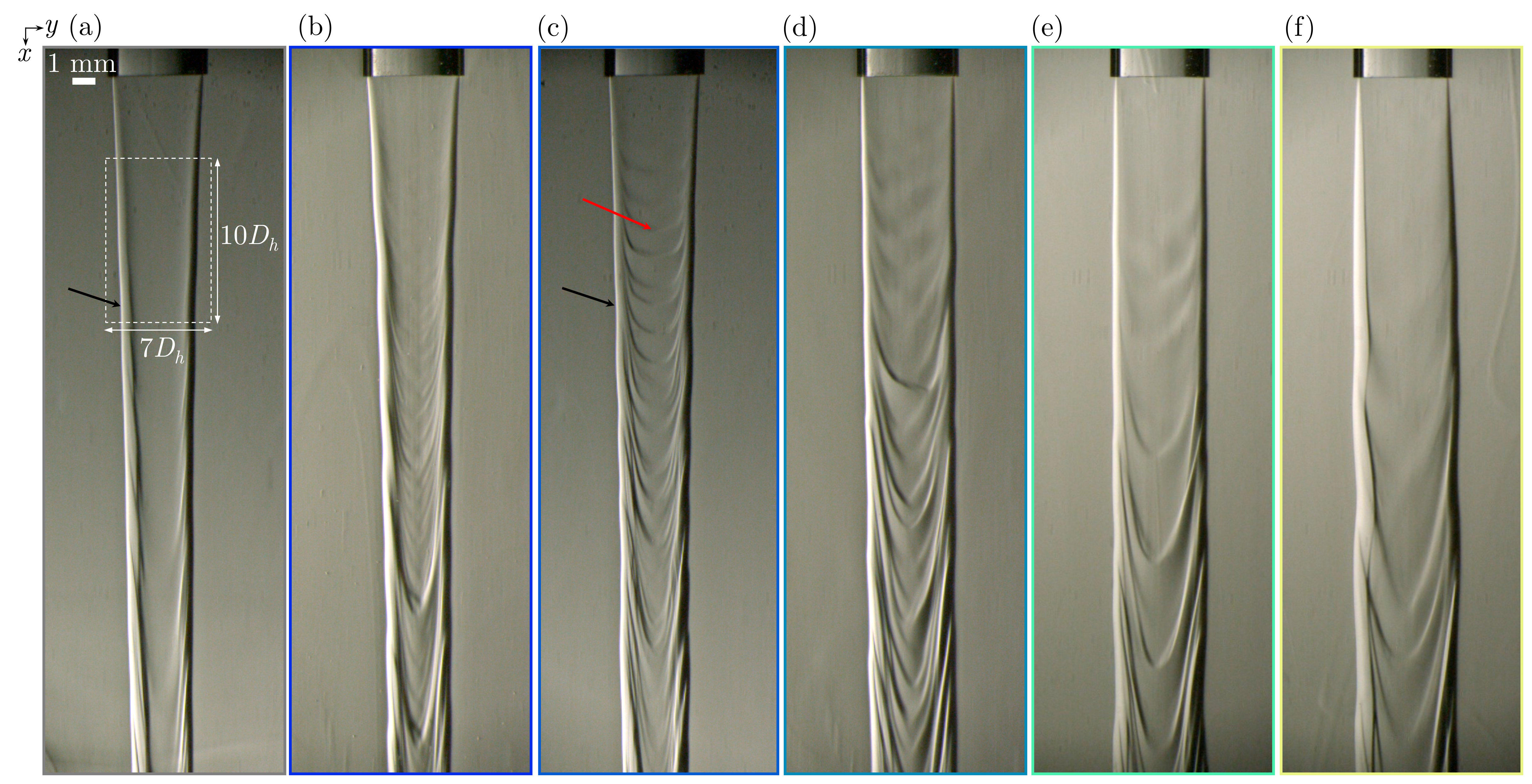}
\caption{Front-view schlieren snapshots of jets at $\textrm{Re} = 100$: (a) Newtonian jet ($\textrm{El} = 0$), (b) $\textrm{El} = 0.004$, (c) $\textrm{El} = 0.013$, (d) $\textrm{El} = 0.041$, (e) $\textrm{El} = 0.066$, and (f) $\textrm{El} = 0.124$. The elasticity number and hence the Weissenberg number $\textrm{Wi} \sim \textrm{El} \cdot \textrm{Re}/(1-\beta)$ increases from left to right. The dashed rectangle indicates the region of interrogation used for the DMD analysis. The black arrows denote the the jet column modes, the red arrow denotes the shear layer modes.}
% \vspace{-14pt}
\label{fig:DMDSnapshots}
\end{figure*}

\noindent DMD decomposes the disturbances into a linearized set of harmonic contributions and identifies the dominant unstable modes. It is a singular value decomposition-based method that can be performed both temporally and spatially \cite{kutz2016dynamic}. To perform spatial DMD in the streamwise direction ($x$-direction), we represent the region of interrogation indicated by the dashed white lines in Fig.\,\ref{fig:DMDSnapshots} as a digital intensity signal $I(x, y, t)$.
% After subtracting the mean of this signal $\bar{I}$ from it, $I(x, y, t) - \bar{I}$
The intensity signal from the high-speed schlieren images can be written as a linear sum of orthogonal modes $\phi(y, t)$, each having a complex amplitude $b$, and a complex wavenumber $k = k_r + ik_i$:  
\begin{equation}
    % I(x, y, t) - \bar{I} = \sum_{j=1}^{N_m} b_j \phi_j(y, t) \exp{(k_r^j + ik_i^j)x}, 
    I(x, y, t) = \sum_{j=1}^{N_x} b_j \phi_j(y, t) \exp{(k_r^j + ik_i^j)x}, 
\end{equation}
where $j=1,2,\,...\,, N_x$ and $N_x$ is the total number of decomposed modes. The magnitude $|b_j|$ determines the relative importance of each mode compared to others, $k_i^j$ is the wavenumber characterizing the periodicity of the mode in the streamwise $x$-direction and $k_r^j$ is the growth rate of the mode where $k_r^j>0$ represents a spatially growing mode.

\noindent The maximum number of resolvable modes is proportional to the length of the data series in space or time (depending on whether spatial or temporal DMD is performed) \cite{schmid2010dynamic}. Retaining all the modes derived from intrinsically noisy experimental data, however, is unwieldy as a large number of relatively weak modes over a wide range of wave numbers or frequencies with approximately equal weighting are observed. \textcolor{black}{To avoid this issue, we identify the optimal number of modes, $N_x^*$, at which the mean squared error between the modal representation and the original schlieren signal exhibits an ``elbow'' corresponding to a distinct change in slope~\cite{kutz2016dynamic}.  To determine the elbow point, we adopt two linear fits, one to the points with mean squared error higher than the estimated $N_x^*$ and one to the points with lower mean square errors. We adjust the estimate of $N_x^*$ until it minimizes the sum of errors of the two fits.} To ensure the robustness of the modes identified by DMD, we also spatially extend the DMD region of interrogation as explained in Appendix\,\ref{app:DMD}. The same primary modes are identified for different regions of interrogation, however, DMD identifies additional slower growing modes when applied on a larger region of interrogation.
% \textcolor{red}{These additional modes represent the onset of nonlinear evolution of the instability that occurs downstream of the jet.}
% Increasing the spatial extent of the interrogation region reveals the same primary modes found in the smaller interrogation region plus additional slower growing modes that develop further downstream of the jet. 

\begin{figure*}[tbhp]
\centering
\includegraphics[width=0.8\textwidth]{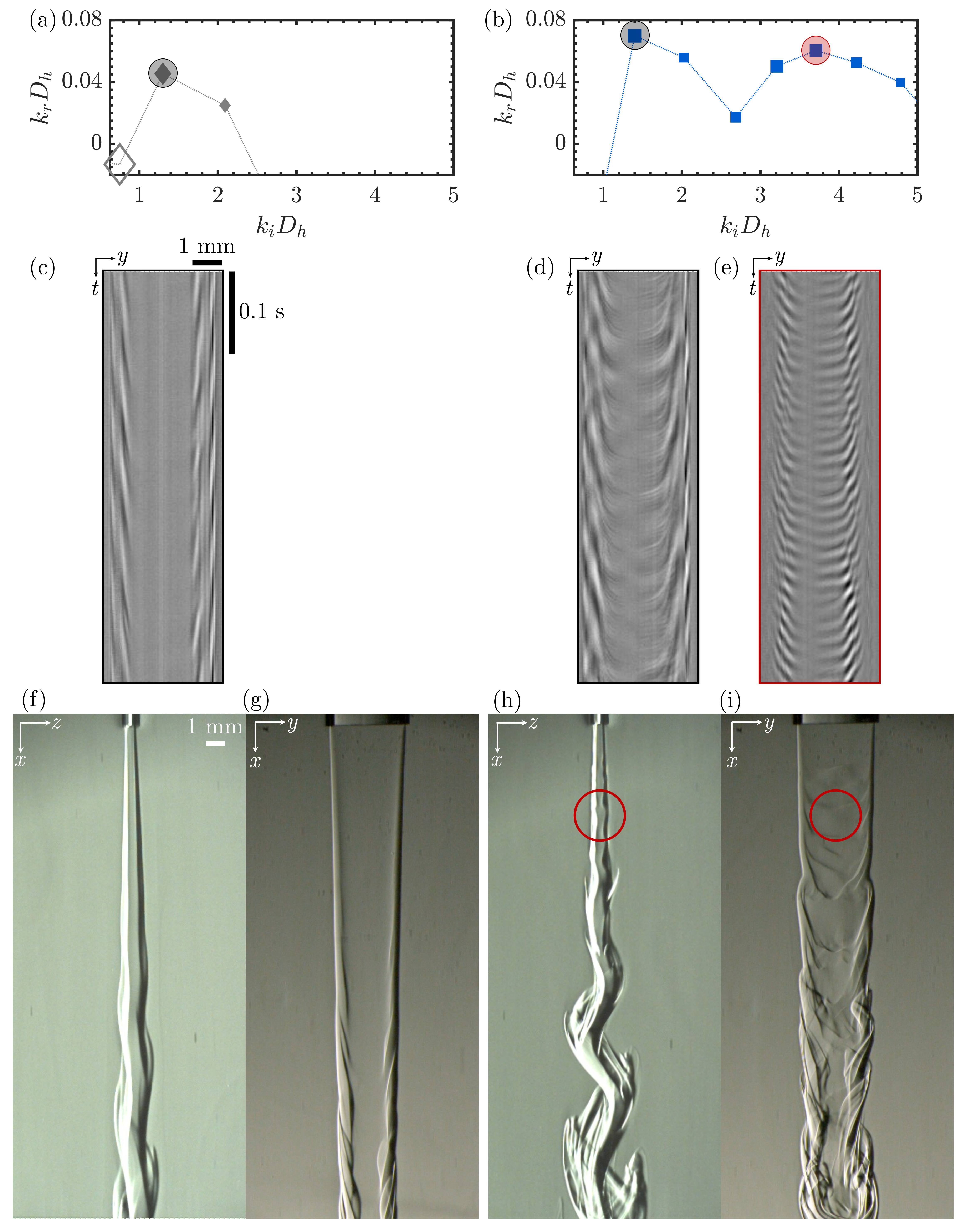}
\caption{ Wavenumber and growth rate of orthogonal modes determined using DMD for (a) a Newtonian and (b) a viscoelastic ($\textrm{El} = 0.013$) jet at $\textrm{Re} = 100$.
% The abscissa and ordinate are nondimensionalized with the jet hydraulic diameter $D_h$.
The abscissa starts from the smallest wavenumber that we can resolve, set by our region of interrogation. The size of the symbols denotes the relative magnitude of the modal amplitude $|b_j|$ with respect to the DC mode. Filled and hollow symbols denote unstable and stable modes, respectively. The dashed lines are guides to the eye for identifying the local maxima in disturbance growth rate. The dominant jet-column and shear-layer modes are marked with gray and red shaded circles, respectively.  (c) The spatio-temporal structure of the orthogonal mode associated with the dominant unstable wavenumber in the Newtonian jet (indicated by the gray circle in (a)). (d) The spatio-temporal structure of the orthogonal mode associated with the dominant unstable low wavenumber in the viscoelastic jet (marked with a gray circle in (b)). (e) The spatio-temporal structure of the orthogonal mode associated with the dominant unstable high wavenumber in the viscoelastic jet (marked with a red circle in (b)). (f, g) Snapshots of the Newtonian jet at $\textrm{Re} = 150$. A long wavelength (small wavenumber) sinuous jet-column mode is visible, but no shear-layer mode is identifiable. (h, i) Snapshots of the viscoelastic jet with $\textrm{El} = 0.013$ at $\textrm{Re} = 150$. Both a low wavenumber sinuous jet-column mode and a high wavenumber shear-layer mode are detected. The shear-layer mode is marked with a red circle.}
% \vspace{-14pt}
\label{fig:DMDModes}
\end{figure*}

\begin{figure*}[b]
\centering
\includegraphics[width=0.5\textwidth]{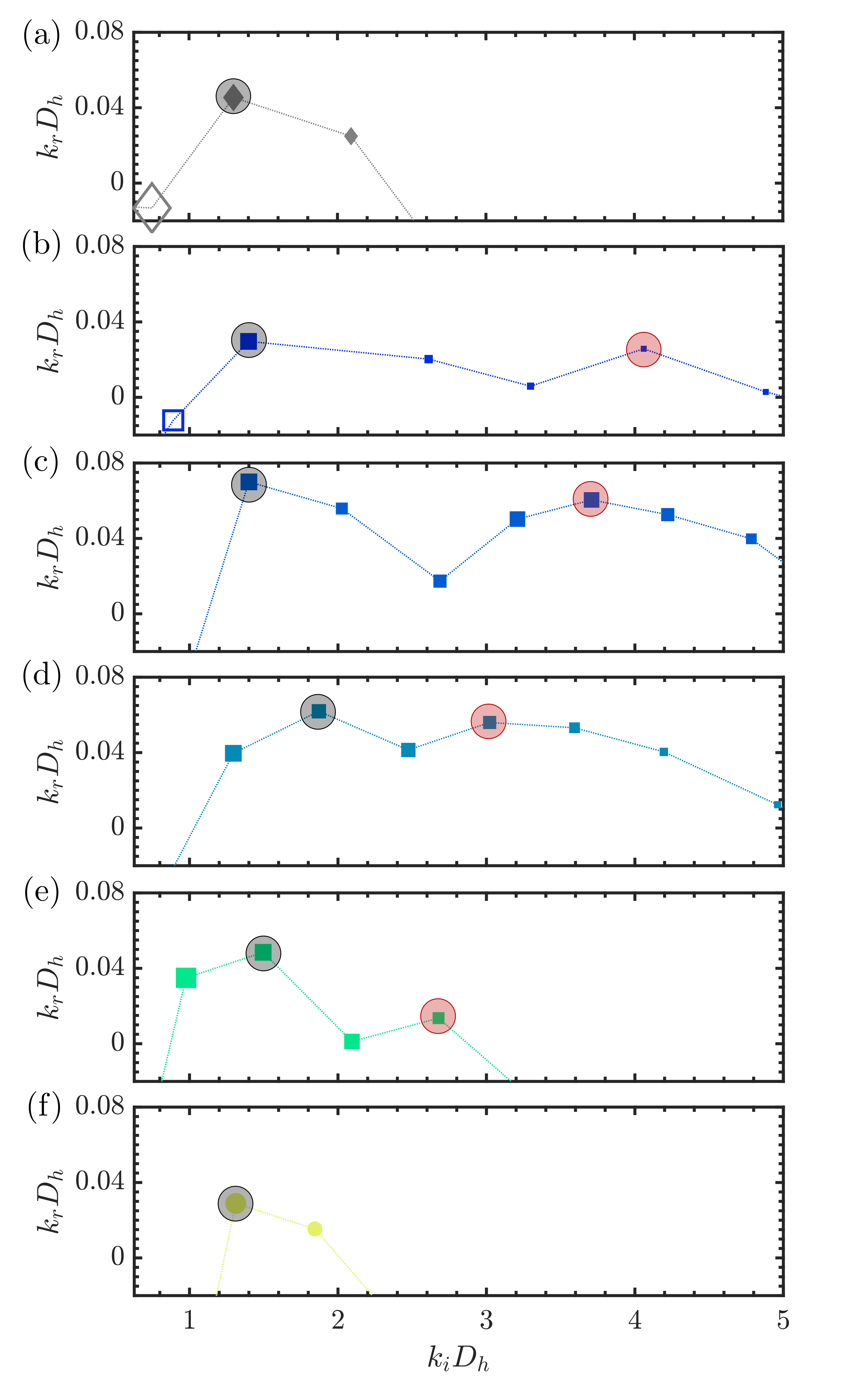}
\caption{ Wavenumber and growth rate of orthogonal modes for (a) Newtonian ($\MyDiamond[draw=water,fill=water]$), (b) $\textrm{El} = 0.004$ (\mysquare{4M50ppm}), (c) $\textrm{El} = 0.013$ (\mysquare{4M100ppm}), (d) $\textrm{El} = 0.0041$ (\mysquare{4M200ppm}), (e) $\textrm{El} = 0.066$ (\mysquare{4M300ppm}), and (f) $\textrm{El} = 0.124$ (\mycircle{8M150ppm}) jets at $\textrm{Re} = 100$. The abscissa and ordinate are nondimensionalized with the jet hydraulic diameter $D_h$. The abscissa starts from the smallest wavenumber that we can resolve based on our region of interrogation. The size of the symbols denotes the relative magnitude of the modal amplitude $|b_j|$. Filled and hollow symbols denote unstable and stable modes, respectively. The jet-column and shear-layer modes with highest growth rate are marked with shaded gray and red circles, respectively. The dashed lines are guides to the eye for identifying the local maxima in growth rate.}
% \vspace{-14pt}
\label{fig:DMDResults}
\end{figure*}

\noindent A temporal DMD analysis (reported in Appendix\,\ref{app:DMD}) shows that the jets are stable in time, indicating that the instabilities are not absolute in nature. A spatial DMD analysis reveals a number of unstable modes indicating the \textit{convective} nature of the instability, as shown in Fig.\,\ref{fig:DMDModes} and \ref{fig:DMDResults}. Each orthogonal mode $\phi_j$ is characterized by three parameters: the wavenumber $k_i^j$, the growth/decay rate $k_r^j$, and the amplitude $|b_j|$. The wavenumber and growth rate of the modes identified for a Newtonian jet and a viscoelastic jet with $\textrm{El} = 0.013$ are shown in Fig.\,\ref{fig:DMDModes}. Here, filled symbols denote spatially growing modes with $k_r >0$, hollow symbols denote spatially decaying modes with $k_r <0$, and the symbol size denotes the relative magnitude of the modal amplitude with respect to the DC mode. The modal amplitude determines the relative contribution of that mode to the reconstruction of the original schlieren image; the larger the amplitude the greater the modal contribution. %Thus, a large filled symbol shows a strongly growing unstable mode and a large hollow symbol shows a strongly decaying stable mode. 
Both the Newtonian and the viscoelastic jet have a dominant unstable mode at a low wavenumber $k_iD_h \simeq 1.3$, while the viscoelastic jet has an additional dominant unstable mode at a higher wavenumber $k_iD_h \simeq 3.7$, as shown in Fig.\,\ref{fig:DMDModes}. %In addition to the difference in the wavenumbers, the orthogonal modes associated with these wavenumbers also show a difference. 
The dominant low wavenumber modes of each jet (Fig.\,\ref{fig:DMDModes}(c) and (d)) denote periodicity at the lateral edges ($y \simeq \pm W/2$) of the jet that results in bulk undulations in the entire jet column. We refer to these modes as ``jet-column'' modes. The additional strong orthogonal mode present in the viscoelastic jet at higher wavenumbers, by contrast, denotes spanwise undulations (clearly observable in Fig.\,\ref{fig:DMDModes}(e)) that are persistent near the center of the jet ($y \sim 0$), which result in a shear-layer instability at the interface between the fast-moving jet and the quiescent background fluid ($z \simeq \pm H/2$). We refer to these modes as ``shear-layer'' modes. To further clarify the difference between jet-column and shear-layer modes, we show side- and front-view snapshots of the Newtonian and viscoelastic jets at $\textrm{Re} = 150$ in Fig.\,\ref{fig:DMDModes}(f)-(i). The Newtonian jet (Fig.\,\ref{fig:DMDModes}(f) and (g)) shows a sinuous jet-column mode, but no shear-layer mode of instability. The viscoelastic jet (Fig.\,\ref{fig:DMDModes}(h) and (g)) similarly shows a sinuous jet-column mode of instability and, additionally, an independent shear-layer instability that grows near the interface between the jet and the background fluid ($z \simeq \pm H/2$ and $-5H \lesssim y \lesssim 5H$, marked with red circles in Fig.\,\ref{fig:DMDModes}(h) and (i)).

\noindent In Fig.\,\ref{fig:DMDResults}, we explore the evolution of the dominant shear-layer mode with elasticity number for the jets shown in Fig.\,\ref{fig:DMDSnapshots}. As the elasticity number increases, the wavenumber associated with the shear-layer mode decreases, while the wavenumber associated with the jet-column mode remains almost unchanged. For $\textrm{El} = 0.124$, we are unable to distinguish two distinct modes of instability and only a single dominant mode is identified. This is consistent with Rallison \& Hinch's argument that the jet-column and shear-layer modes merge at higher elasticity numbers \cite{Rallison1995}. These findings allow us to rationalize the observations in Fig.\,\ref{fig:DMDSnapshots}; the emergence of the viscoelastic shear-layer instability destabilizes the jet, whereas its subsequent merging with the jet-column mode partially  re-stabilizes the jet, accounting for the competing effects of increasing elasticity number on jet stability. Our results are, to our knowledge, the first experimental manifestation of the shear-layer and jet-column modes in viscoelastic jets that were previously identified in a temporal linear stability analysis \cite{Rallison1995}. The shear-layer mode present for small non-zero elasticity numbers within the shear layer at the edge of the jet is independent of the jet geometry and velocity profile and orthogonal to the longer wave jet-column instability modes. In this low El regime, the effects of elasticity are important only within a thin sheared layer close to the jet edge where the fluid inertia is smaller (compared to the core of the jet), but where the shear rate is higher compared to the core of the jet, resulting in a large first normal stress difference.
% (the normal stresses difference depends quadratically on shear rate).
Rallison \& Hinch argue that the interplay between inertial and elastic effects %is captured in the axial momentum equation along a streamline and 
results in a low pressure region just outside the jet and a crowding of the streamlines. Such crowding induces a dilation of the streamlines just inside the jet in the thin edge region where elasticity dominates. The crowding and dilation of the streamlines generates the local shear-layer instability. Their calculations show that the wavenumber associated with the local shear-layer instability decreases with an increase in elasticity number \cite{Rallison1995}, which is supported by our experimental findings. The confinement of the shear-layer mode of instability to the very edge of the jet ($z = \pm H/2$) implies that it can occur for both planar and axisymmetric jets, irrespective of whether the fluid jet column is unstable to a sinuous or a varicose mode. As the elasticity number increases, the thickness of the sheared fluid layer in which elastic effects are dominant increases, which eventually results in the merging of the shear-layer instability and the jet-column instability \cite{Rallison1995}. \textcolor{black}{While the shear-layer instability identified in this study is an inertio-elastic instability that is limited to the viscoelastic jets, a Newtonian jet can also undergo an inertial shear-layer instability but typically at much higher Reynolds numbers ($\textrm{Re} \geq 8000$) \cite{thomas1991experimental}. For the range of Reynolds numbers investigated, we do not observe an inertial and spatially growing shear-layer instability in the Newtonian jet.}

\noindent The jet instabilities demonstrated here lead ultimately to the transition to turbulence. The emergence of the spatially growing viscoelastic shear-layer mode at small, but finite, elasticity numbers induces a transition to turbulence at a lower Reynolds number compared to the Newtonian jet, as shown in Appendix\,\ref{app:snapshots}. At a fixed Reynolds number, the jets for which a viscoelastic shear-layer mode emerges show a transition to turbulence at a streamwise Eulerian location that is closer to the nozzle compared to the Newtonian jet. The merging of the shear-layer mode with the jet-column mode at higher elasticity numbers results in a delay in transition to turbulence until higher Reynolds number (compared to the low elasticity number jets). Similarly, at a fixed Reynolds number, our schlieren imaging shows that the high elasticity number jets transition to turbulence at streamwise Eulerian locations that are further from the nozzle compared to the low elasticity number jets. This effect is explored and quantitatively studied for turbulent planar jets in the next section.

\subsection{Self-similarity of planar jets}
\label{sec:similarity}

\begin{figure*}[tbhp]
\centering
\includegraphics[width=0.99\textwidth]{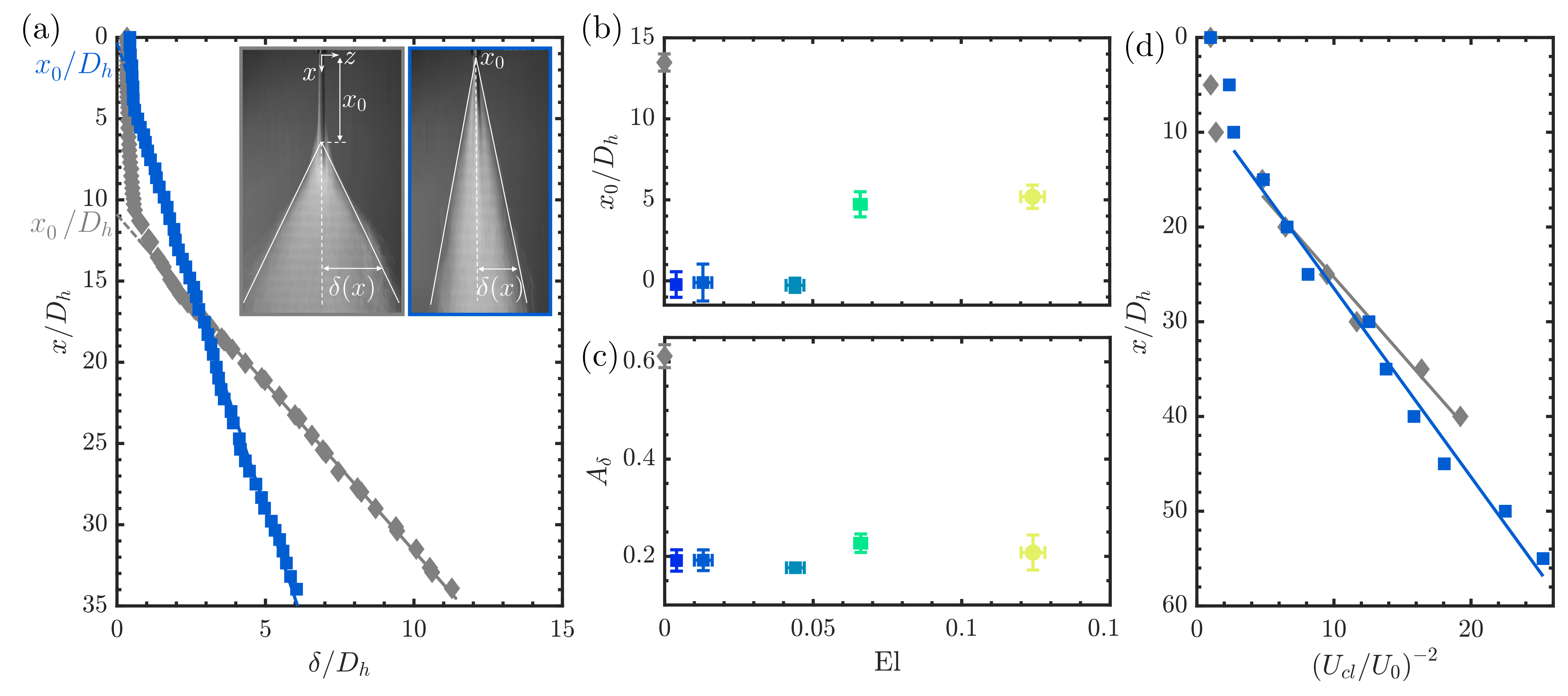}
\caption{(a) Turbulent spreading of a Newtonian jet ($\MyDiamond[draw=water,fill=water]$) and a viscoelastic jet with $\textrm{El} = 0.013$ (\mysquare{4M100ppm}) calculated from the standard deviation of intensity fluctuations in side-view schlieren images, shown in the inset, at $\textrm{Re} = 400$. To enable identification of the nozzle location and the laminar region of the jets in the insets, the standard deviation images are overlaid on the image of the mean of each pixel over the duration of the experiment. The lateral spreading of the jets
% is the distance of the jet boundary from the jet centerline 
at dimensionless streamwise locations $x/D_h$ is denoted by $\delta$. The onset of the self-similar turbulent region correspond to a virtual origin $x_0$. The symbols denote the profile extracted from the images, the solid lines denote a linear fit to the turbulent region. The relationships between the jet spreading and the streamwise distance from the nozzle in the turbulent region are $\delta/D_h = (0.61\pm 0.02)(x-x_0)/D_h$ and $\delta/D_h = (0.19\pm 0.02)(x-x_0)/D_h$ for the Newtonian and viscoelastic jets, respectively. (b) The extrapolated origin $x_0$ of turbulent jets as a function of elasticity number $\textrm{El}$ at $\textrm{Re} = 400$. %The horizontal error bars show the error of elasticity number calculated based on error propagation principles and shown in Table\,\ref{tab:propertySI}. The vertical error bars show the standard deviation of $x_0$ calculated for the left side and right side of the jet for two experiments. 
(c) Coefficient of spreading $A_{\delta}$ versus elasticity number at $\textrm{Re} = 400$. %The vertical error bars show the standard deviation of $A_{\delta}$ calculated for the left side and right side of the jet for two experiments. 
(d) The normalized streamwise mean centerline velocity, $U_{cl}$/$U_0$, for a Newtonian and a viscoelastic ($\textrm{El} = 0.013$) jet scales as $U_{cl}/U_0 \sim (x/D_h)^{-1/2}$, where $U_0$ is the mean velocity at the nozzle exit. The relationships between the normalized velocity and the streamwise distance from the nozzle in the turbulent region are $\left(U_{cl}/U_0\right)^{-2} = (0.60\pm 0.06)(x-x_0)/D_h$ and $\left(U_{cl}/U_0\right)^{-2} = (0.50\pm 0.04)(x-x_0)/D_h$ for the Newtonian and viscoelastic jets, respectively.% Fan and velocity profile of Newtonian and viscoelastic jets in the streamwise direction $x$: (a) Streamwise spreading, \textit{i.e.}, fan, of turbulent Newtonian and viscoelastic jets ($M_w = 4\times 10^6 \textrm{ g/mol},~ c=100\textrm{ ppm}$) calculated from the standard deviation of side view schlieren images, shown in the inset, at $\textrm{Re = 400}$. The fan of the jets
% % is the distance of the jet boundary from the jet centerline 
% at dimensionless streamwise locations $x/D_h$, is denoted by $\delta_w(x)$ and $\delta_p(x)$ for the Newtonian and viscoelastic jets, respectively. The start of the jet turbulent region is marked with $x_0^w$ and $x_0^p$ for the Newtonian and viscoelastic jets, respectively. The symbols show the profile extracted from the inset images and the solid lines show a linear fit to where the jets are turbulent. (b) Streamwise mean centerline velocity, $U_cl$, for the Newtonian and viscoelastic jets. The velocities are non-dimensionalized with the mean velocity at the nozzle exit, $U_0$. The symbols show the LDV measurements. To better represent the $U_cl \sim x^{-1/2}$ scaling, streamwise values of $\left(U_{cl}/U_0\right)^{-2}$ are plotted and the solid lines show linear fits to them. The inset shows the streamwise profile of $U_{cl}/U_0$.
}
% \vspace{-14pt}
\label{fig:Selfsimilarity}
\end{figure*}

\noindent The nonlinear evolution of the shear-layer and jet-column instability modes results in the transition of the viscoelastic jets to an EIT state. %At the EIT state, turbulent jets show self-similarities in both their time-averaged and spectral properties. In this section, we focus on understanding this self-similar behavior at the EIT state for viscoelastic planar jets and the impact of the identified instabilities on this behavior. 
In this state, the jets spread laterally, and the spreading is a measure of the jet entrainment process. To quantify this lateral spreading, we calculate the standard deviation of the time\textendash varying intensity for each pixel in the side-view schlieren images over the duration of the experiment, as shown in the inset of Fig.\,\ref{fig:Selfsimilarity}(a). We employ Otsu's method \cite{otsu1979threshold, huang2009optimal, nolan2013conditional} to detect the lateral boundary of the region affected by turbulent fluctuations. Following \cite{miller1996measurements,guimaraes2020direct}, we define the jet spreading parameter $\delta$ as
% the jet half\textendash width, \textit{i.e.},
the distance from the centerline of the jet to the boundary, as reported in Fig.\,\ref{fig:Selfsimilarity}(a) for jets at $\textrm{Re} = 400$. 
% The streamwise distance from the nozzle and the jet spreading are nondimensionalized with the jet hydraulic diameter $D_h$. 
%To enable identification of the nozzle location and the laminar region of the jets, the standard deviation images are overlaid on the image of the mean of each pixel over the duration of the experiment. 
For both Newtonian and viscoelastic jets, we find the jet spreading can be written in the form $\delta/D_h  = A_{\delta} \left(x - x_{0}\right)/D_h$, where $A_{\delta}$ is the coefficient of spreading and $x_0$ denotes the extrapolated or virtual origin of the turbulent jet. While the $\delta \sim x$ scaling is consistent with the self-similarity analysis \cite{guimaraes2020direct}, such a scaling analysis does not provide values for $A_{\delta}$ and $x_0$ and experiments or numerical simulations are required to obtain them. \textcolor{black}{For the Newtonian jet we find a coefficient of spreading that is larger than the typical values reported in the literature at Reynolds numbers that are higher than our investigated Reynolds number \cite{kotsovinos1976note}. Our results are in qualitative agreement with \cite{suresh2008reynolds}, a study at Reynolds numbers comparable to ours. A quantitative comparison of $A_{\delta}$, however, is challenging as our calculation method for $\delta$ is different from that used in \cite{suresh2008reynolds}.}
% Both jets thus spread self-similarly, as evidenced by the scaling $\delta \sim x$ \cite{guimaraes2020direct}.
Our experiments show that all viscoelastic jets transition to turbulence at an Eulerian position that is closer to the nozzle compared to the Newtonian jet resulting in dramatically reduced values of the virtual origin $x_0$, as shown in Fig.\,\ref{fig:Selfsimilarity}(b).  For small, non-zero elasticity numbers ($0.004 \lesssim \textrm{El} \lesssim 0.04$), the transition to turbulence occurs at a streamwise location very close to the nozzle, consistent with the emergence of the shear-layer instability in these jets. A further increase in elasticity number to values where the shear-layer mode merges with the jet-column mode ($\textrm{El} \gtrsim 0.04$) results in a moderate increase in $x_0/D_h$, as shown in Fig.\,\ref{fig:Selfsimilarity}(b). The coefficient of spreading is smaller for viscoelastic jets compared to the Newtonian jet, but is unchanged within our experimental resolution for a range of elasticity numbers $0.004<\textrm{El}<0.124$, indicating that viscoelastic jets in the EIT state maintain a universal entrainment rate. 
% The smaller spreading of the viscoelastic jet results in a lower rate of entrainment and a smaller rate of vorticity diffusion. From conservation of mass, this implies that the rate of decay of the centerline velocity is smaller than that of the Newtonian jet. %consistent with the universality of the EIT state \cite{yamani2021spectral}. 

\noindent Furthermore, we measure the mean centerline velocity $U_{cl}$ for the Newtonian and viscoelastic ($\textrm{El} = 0.013$) jets at fixed Eulerian positions on the centerline of the jet and in the streamwise direction using LDV. The center of the probe volume is aligned with the centerline of the jet, and the probe volume is smaller than the smallest lateral dimension of the jet enabling good resolution of the centerline velocity. %However, the jet velocity profile is approximately parabolic at the measurement positions close to the nozzle resulting in an increase in the standard deviation of the velocity measurements close to the nozzle. \noindent %The mean centerline velocity is nondiemensionalized with the mean velocity at the nozzle, $U_0$. 
The velocity profile of both jets follows $U_{cl} \sim x^{-1/2}$, which suggests that the velocity profile in the turbulent region of the jets can be written in the similarity form  $\left(U_{cl}/U_0\right)^{-2} = A_{U}\left(x - x_0\right)/D_h$ \cite{guimaraes2020direct}, as shown in Fig.\,\ref{fig:Selfsimilarity}(d), where $A_{U}$ is the similarity coefficient for the velocity profile. \textcolor{black}{For the Newtonian jet, the velocity similarity coefficient matches the value reported in an earlier study \cite{suresh2008reynolds} at the same Reynolds number. The viscoelastic jet has a smaller similarity coefficient, $A_{U}$, compared to its Newtonian counterpart, which is consistent with its smaller coefficient of spreading, $A_{\delta}$. The smaller rate of spreading of the viscoelastic jet results in a lower rate of entrainment and a smaller rate of vorticity diffusion. From conservation of mass, this implies that the rate of decay of the centerline velocity is slower than that of the Newtonian jet.} 
% The identified self-similar jet characteristics in our experiments are in agreement with self-similarity analysis of planar jets \cite{guimaraes2020direct}.

% \noindent The identified self-similar jet characteristics in our experiments are in agreement with self-similarity analysis of planar jets \cite{guimaraes2020direct}. Our study on the impact of El on the streamwise distance for the onset of turbulence    

% Self-similarity of the time-averaged properties in turbulent free-shear flows for Newtonian fluids have been studied in the past and a comprehensive table summarizes the scaling relations derived from these self-similarities in chapter 12 of  \cite{Kundu2016fluid}. It was only recently that the self-similarity of viscoelastic turbulent jets was explored \cite{guimaraes2020direct}. This self-similarity analysis provides scaling relations for shear layer thickness, \textit{i.e.}, spreading of the jet,  centerline velocity, and the maximum polymer stresses for a submerged planar jet of viscoelastic fluid independent of the constitutive model used for the polymer stress tensor in the Cauchy momentum equation. Here we focus on the self-similarity of the jet spreading and the centerline velocity for the viscoelastic jet at the EIT state and its differences with self-similarity of the Newtonian jet at NT.
% ($M_w = 4\times 10^6 \textrm{ g/mol},~ c=100\textrm{ ppm},~ \textrm{El} = 0.013$) at $\textrm{Re} = 400$. 

\subsection{Lagrangian coherent structures in elasto-inertial turbulence of planar jets}\label{sec:ResultsLCS}

\noindent The concept of Lagrangian coherent structures can be extended from Newtonian turbulence to elasto-inertial turbulence to understand how viscoelasticity affects the evolution of material regions in a time-dependent turbulent flow \cite{dubief2022first}. Coherent structures are local material regions of the flow whose dynamics contain a significant portion of the total energy in the flow and that remain correlated over sufficiently large spatio-temporal scales \cite{dubief2022first}. Common experimental techniques such as LDV or particle image velocimetry must be coupled with advanced non-linear dynamics data processing techniques such as the Finite Time Lyapunov Exponent \cite{Shadden2011} or Diffusion Barrier Strength approach \cite{Haller2018} to determine the spatio-temporal evolution of material regions. Schlieren imaging directly reveals the Lagrangian evolution in the boundaries of the fine-scale material regions as they are advected through the domain.

\begin{figure*}[tbhp]
\centering
\includegraphics[width=\textwidth]{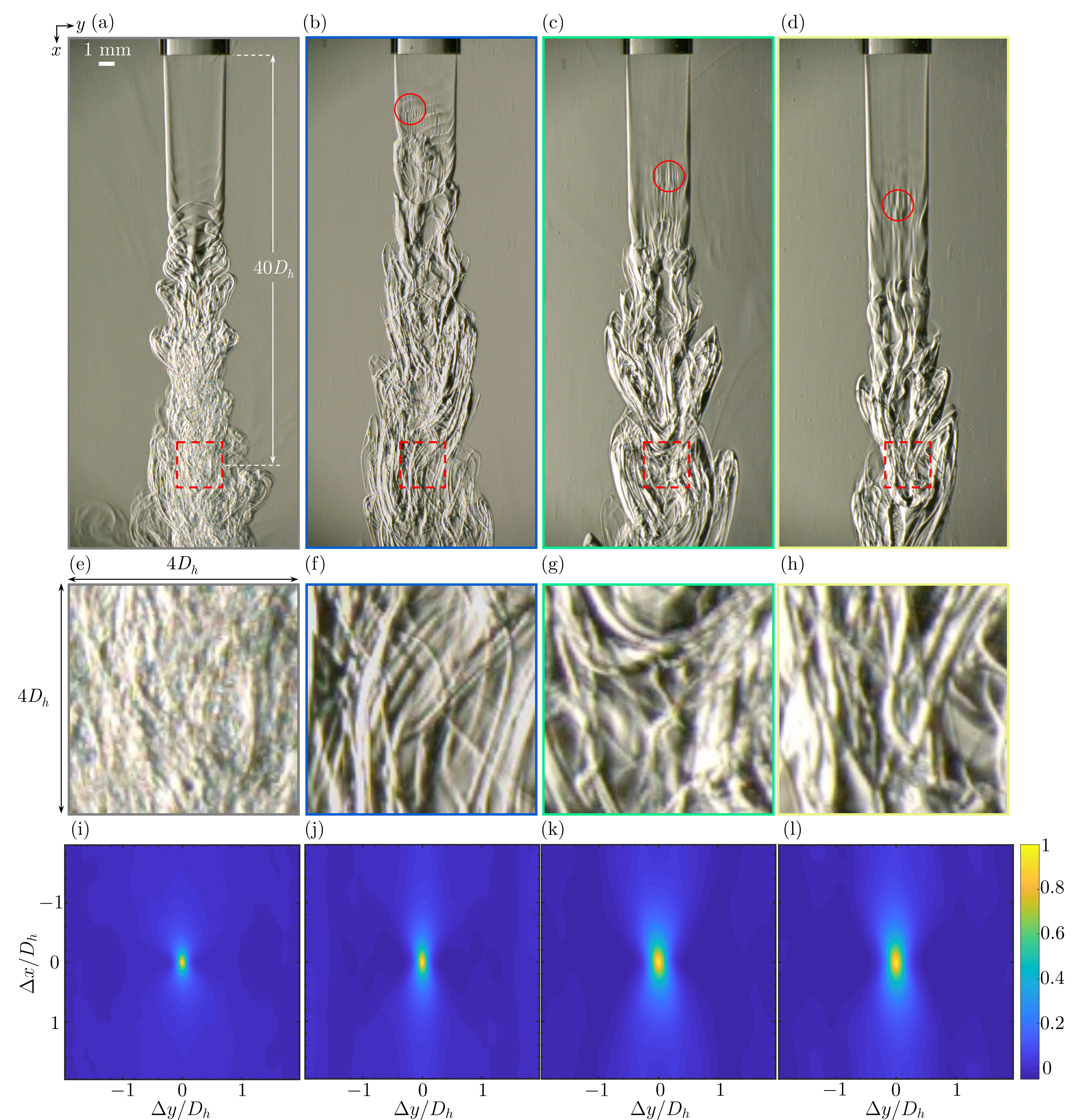}
\caption{Front-view schlieren snapshots of (a) a Newtonian jet and (b-d) viscoelastic jets with (b) $\textrm{El} = 0.013$ (c) $\textrm{El} = 0.066$, and (d) $\textrm{El} = 0.124$ at $\textrm{Re} = 400$. The regions of interrogation of size $4D_h$ are indicated by the red squares. (e-f) Magnified regions of interrogation for the jets shown in (a-d), highlighting how an increase in elasticity number modifies the structure of turbulence. (i-l) Two dimensional auto-covariance of the regions of interrogation shown in (e-f). The color bar indicates the average value of the 2D autocovariance $\bar{R}$.}
% \vspace{-14pt}
\label{fig:SpatialImages}
\end{figure*}

\noindent Our schlieren images provide unique visualization of the characteristic features in fully-developed elasto-inertial turbulence and Newtonian turbulence in planar jets, as shown in Fig.\,\ref{fig:SpatialImages}(a)-(d) and Supplemental Material, Movie\,S8 \cite{supp} for $\textrm{Re} = 400$. Similar to the behavior observed at lower Reynolds numbers, increasing the elasticity number first has a strongly destabilizing effect and makes the jet turbulent at a streamwise location closer to the nozzle, while a further increase in elasticity number moves the onset of the turbulent region further downstream due to merging of shear-layer and jet-column modes, as shown in Fig.\,\ref{fig:SpatialImages}(a)-(d). The shear-layer modes create streamwise aligned coherent structures that are an elasto-inertial analog reminiscent of Emmons turbulent spots \cite{emmons1951laminar,cantwell1978structure,marxen_jfm2019}, marked with red circles in Fig.\,\ref{fig:SpatialImages}(b)-(d). 
% and were first visualized for Newtonian turbulence in 1978 \cite{cantwell1978structure} 
The absence of shear-layer modes in the Newtonian jet prevents observation of these local turbulent patches, where the jet transitions to turbulence before turbulent spots can form due to the braiding and roll-up of vortices arising from the nonlinear evolution of the jet-column modes.

\begin{figure*}[h]
\centering
\includegraphics[width=0.99\textwidth]{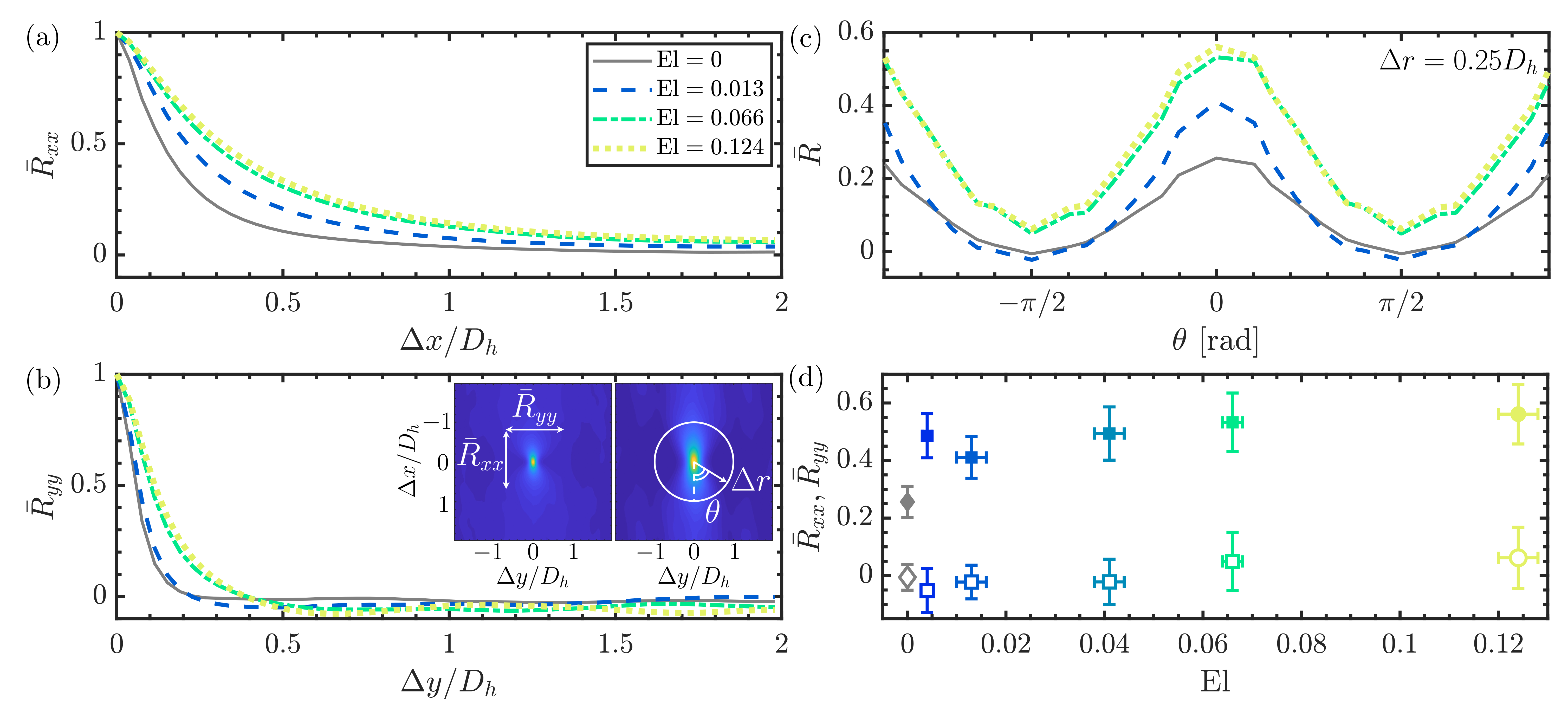}
\caption{(a) Decay in the streamwise ($\bar{R}_{xx}$) and (b) spanwise ($\bar{R}_{yy}$) autocovariance versus nondimensional streamwise ($\Delta x/D_h$) and spanwise ($\Delta y/D_h$) displacements for a Newtonian ($\textrm{El} = 0$) and for viscoelastic jets with $\textrm{El} = 0.013$, $\textrm{El} = 0.066$, and $\textrm{El} = 0.124$ at $\textrm{Re} = 400$. The left inset in (b) defines $\bar{R}_{xx}$ and $\bar{R}_{yy}$, the right inset defines the polar coordinate system. (c) Autocovariance in a polar coordinate system ($R\left(\Delta r, \theta\right)$) for $\Delta r = 0.25D_h$. (d) Streamwise (filled symbols) and spanwise (hollow symbols) values of the autocovariance at $\Delta r = 0.25D_h$ as a function of elasticity number. The horizontal error bars denote the uncertainty in the elasticity number shown in Table\,\ref{tab:propertySI}. The vertical error bars denote the standard deviations of $\bar{R}_{xx}$ and $\bar{R}_{yy}$ at $\Delta r = 0.25D_h$. % (b-d) Autocovariance of the studied planar jets in a polar coordinate system ($R\left(\Delta r, \theta\right)$) for (b) $\Delta r = 0.25D_h$, (c) $\Delta r = 0.5D_h$,  and (d) $\Delta r = D_h$. The inset in (b) shows the polar coordinate system for the two dimensional autocovariance for planar jet of PEO solution with $\textrm{El} = 0.013$ ($M_w = 4\times 10^6 \textrm{ g/mol},~ c=100\textrm{ ppm}$).
}
% \vspace{-14pt}
\label{fig:SpatialResults}
\end{figure*}

\noindent To identify the key features of these Lagrangian coherent structures (LCSs), we select a square region of interrogation in the front-view schlieren images of size $4D_h \times 4D_h$, with its centroid located at the streamwise distance of $x = 40D_h$ from the nozzle. %This region of interrogation has more than $1D_h$ lateral distance from the boundary of the jet and is considered homogeneous due to the large aspect ratio of the jet. To evaluate the robustness of this assumption, a smaller region of interrogation with side $3D_h$ has been studied and the center of the interrogation region was moved in the streamwise direction to $x = 35D_h$. All the investigated regions showed similar results to the original region of interrogation, ensuring us that our analysis is carried out at a region of interrogation where homogeneous turbulence exists.  
We evaluate the size and orientation of the turbulent structures by calculating the two-dimensional (2D) autocovariance for the region of interrogation for each jet at a fixed value of $\textrm{Re} = 400$, as shown in Fig.\,\ref{fig:SpatialImages}(e)-(h). The autocovariance is first calculated for each schlieren snapshot,

\begin{equation}
    R(\Delta x, \Delta y, t) = \frac{\sum_{x,y}\left(I(x,y,t) - \bar{I}\right)\left(I\left(x+\Delta x, y+\Delta y , t\right)-\bar{I}\right)}{\left[\sum_{x,y}\left(I(x,y,t) - \bar{I}\right)^2\sum_{x,y}\left(I\left(x+\Delta x , y +\Delta y , t\right) - \bar{I}\right)^2\right]^{1/2}},
\end{equation}
where $I(x,y,t)$ is the intensity of the schlieren image at position $(x,y)$ and at time $t$, and $\bar{I}$ is the average intensity of the region of interrogation over space and time. The average value of the 2D autocovariance is then calculated by computing the expectation over all snapshots, $\bar{R}(\Delta x, \Delta y) = \mathbf{E}\left[R\left(\Delta x, \Delta y, t\right)\right]$. The contours of $\bar{R}$ are shown in Fig.\, \ref{fig:SpatialImages}(i)-(l), where the color bar denotes autocovariance values of $-0.05<\bar{R}<1$ from dark blue to yellow. Increasing elasticity number increases $\bar{R}$ in the streamwise direction, $x$, as shown in  Fig.\,\ref{fig:SpatialResults}(a) and (b), where we report the streamwise and spanwise values of the autocovariance, $\bar{R}_{xx}$ and $\bar{R}_{yy}$, (see inset of Fig.\,\ref{fig:SpatialResults}(b) for definition) as a function of nondimensionalized streamwise and spanwise displacements, $\Delta x/D_h$ and $\Delta y/D_h$, respectively. The streamwise correlation decays more slowly than the spanwise correlation for all jets, due to the non-zero mean shear in the flow. In addition, the viscoelastic jets have a higher streamwise correlation than the Newtonian jet for all values of El. The rate of decay in the spanwise correlation, however, remains similar for Newtonian and viscoelastic jets. Elasto-inertial turbulence is thus characterized by coherent structures in the form of material regions that are more elongated in the streamwise direction compared to those observed in Newtonian turbulence. 
% Further increase in the elasticity number ($\textrm{El} = 0.066$, and 0.124) maintains the streamwise correlation for the viscoelastic jets significantly higher than that of the Newtonian jet. 
% increases both the streamwise correlation and spanwise correlations simultaneously, showing that not only material elements become more elongated in the streamwise direction but they also become larger in the spanwise direction. 
This is further confirmed by a polar representation of the 2D autocovariance, as shown in Fig.\,\ref{fig:SpatialResults}(c) for a radial displacement $\Delta r = 0.25D_h$. In this representation, $\theta = 0$ and $\theta = \pm \pi/2$ represent $\bar{R}_{xx}$ and $\bar{R}_{yy}$, respectively.  The variations with elasticity number in the values of streamwise and spanwise correlations at $\Delta x/D_h = \Delta y/D_h = 0.25D_h$ are summarized in Fig.\,\ref{fig:SpatialResults}(d). The viscoelastic jets have a streamwise correlation value that is statistically higher than that of the Newtonian jet for all elasticity numbers, while the spanwise correlations of the jets do not show a statistically significant difference. The observation that the streamwise and spanwise correlations are independent of elasticity number for the viscoelastic jets further supports the universality of EIT \cite{yamani2021spectral}. The schlieren observations of elongated and flow-aligned material domains are consistent with the suppression of streamwise vortices by EIT that has previously been reported in numerical simulations of turbulent wall-bounded shear flows \cite{stone2004polymer, li2007polymer,xi2010turbulent, Samanta2013, dubief2013mechanism}.

% These coherent structures grow in size in both streamwise and spanwise direction as elasticity number increases.
% Furthermore, the results of this Lagrangian spatial analysis is in agreement with our results of the turbulent kinetic energy power spectral density calculated from the streamwise velocity fluctuations. The higher power in EIT compared to NT at smaller frequencies, or equivalently at smaller wavenumbers due to the validity of the frozen flow hypothesis, is consistent with the schlieren observation of larger LCSs in the streamwise direction for EIT compared to NT. 

\noindent \textcolor{black}{Finally, it is important to note that the global Weissenberg number defined based on the bulk flow properties does not characterize the dynamics of the viscoelastic jets in the fully turbulent region studied above. Here we define a local Weissenberg number based on turbulent properties as $\textrm{Wi}_{T} = \lambda u_{cl}^{RMS} / \ell_T$, where $\ell_T$ is the Taylor microscale. Adopting a frozen flow hypothesis \cite{taylor1938spectrum}, the Taylor microscale is defined as \cite{pope2000turbulent},
\begin{equation}
    \ell_T = \left[-\frac{1}{2} \frac{d^2 \bar{R}_{xx}}{d(\Delta x/D_h)^2}\right]^{-1/2},
\end{equation}
such that the parabolic curve fitting of the computed $\bar{R}_{xx}$ at $\Delta x/D_h = 0$ has the form,  
\begin{equation}
    p(\Delta x/D_h) = 1 - \frac{(\Delta x/D_h)^2}{(\ell_T/D_h)^2},
\end{equation}
and therefore $p(\ell_T) = 0$. Based on the parabola osculating $\bar{R}_{xx}$ of the viscoelastic jet with $\textrm{El} = 0.013$, shown in Fig.\,9(a), the Taylor microscale is $\ell_T = 0.183D_h = 1.33 \times 10^{-4} \textrm{ m}$, resulting in $\textrm{Wi}_{T} = 4.09$. 
% In addition, we can define a local Deborah number which compares the relaxation timescale with $\tau_{T} = \ell_{T}/U_{cl}$, which thus has the definition $\textrm{De}_{T} = \lambda/\tau_{T} = 14.51$.
This local Weissenberg number is greater than unity, which shows the relative importance of elasticity. This information together with the dimensionless parameters defined based on the bulk properties of the jets will allow for data-infused simulations of the viscoelastic planar jets.
% This information, together with the dimensionless parameters defined based on the bulk properties of the jets, will allow for data-infused simulations of the viscoelastic planar jets.
}

% Also note that FENE constitutive equation is commonly used for direct numerical simulations of turbulent viscoelastic flows. The comparison of both global and local Weissenberg numbers calculated above with the finite extensibility parameter, $L_{max}$, can show the degree of nonlinearity for these simulations \cite{yamani2022master}.

\subsection{Power spectral density of elasto-inertial turbulence}
\label{sec:spectrum}
\begin{figure*}[tbhp]
\centering
\includegraphics[width=0.99\textwidth]{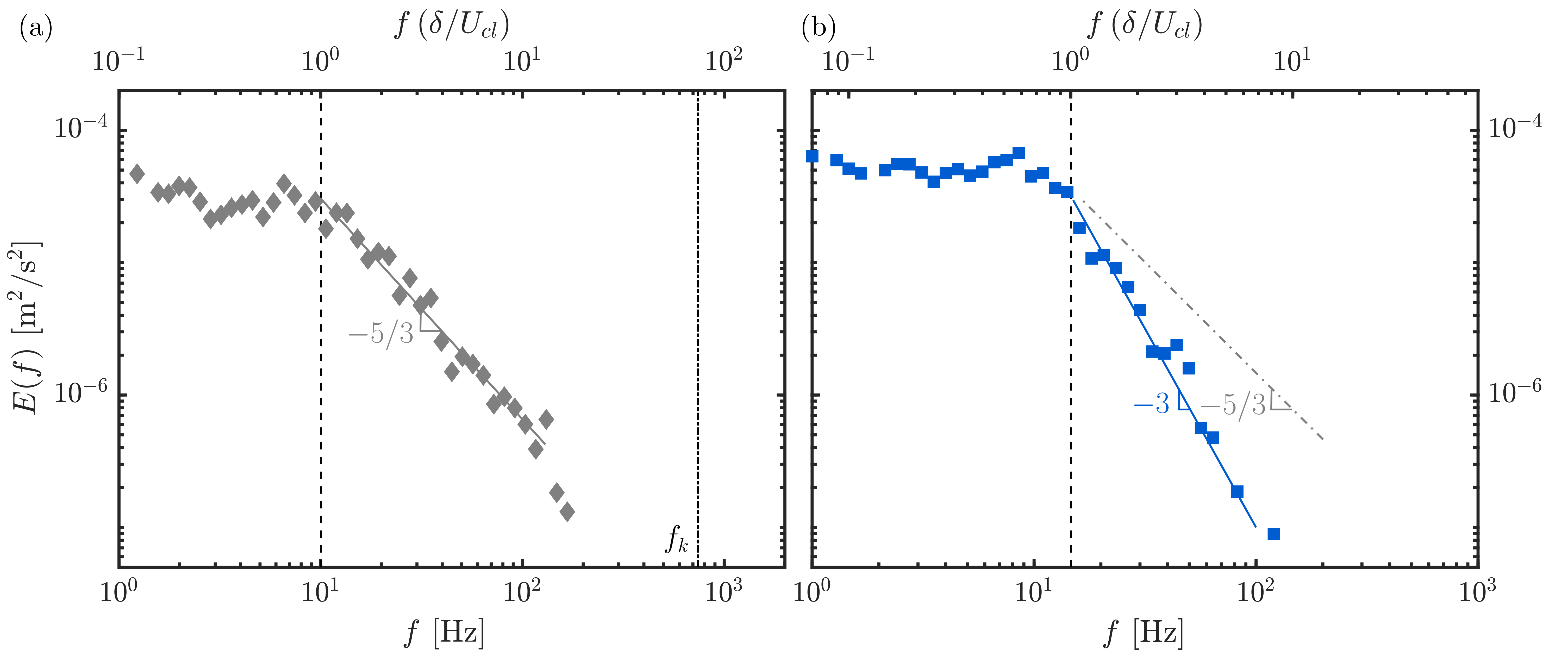}
\caption{Turbulent kinetic energy PSD for (a) a Newtonian jet ($\MyDiamond[draw=water,fill=water]$) and (b) a viscoelastic jet (\mysquare{4M100ppm}) with $\textrm{El} = 0.013$ at $\textrm{Re} = 400$, calculated at a streamwise distance $x = 40D_h$ from the nozzle and at the centerline of the turbulent jet. The bottom abscissa is the dimensional frequency, the top abscissa is the nondimensional frequency, normalized by $\delta/U_{cl}$, where $\delta$ is the spreading of the jet and $U_{cl}$ is the mean centerline velocity at $x = 40D_h$. The dimensionless frequency $f\left(\delta/U_{cl}\right) = 1$ denoting the onset of the inertial and elasto-inertial range of the spectrum is indicated by dashed lines. The Kolmogorov frequency $f_k$, at which the dominance of viscous effects terminates the inertial range of the spectrum, is indicated by the dashed-dotted line.}
% \vspace{-14pt}
\label{fig:PSD}
\end{figure*}

% In addition to the self-similarity in the time-averaged properties of the viscoelastic planar jets discussed in Sec.\,\ref{sec:similarity}, we also identify self-similar behavior in the spectral properties of their turbulent fluctuations.
\noindent  We calculate the PSD of the velocity fluctuations $E(f)$, also referred to as the turbulent kinetic energy spectrum, at a fixed Eulerian location $x = 40D_h$ and at the centerline of the jet from the fluctuating velocity time series $u_{cl}(t)$. For the Newtonian jet, the PSD exhibits a power-law decay with exponent $-1.7 \pm 0.2$ in the inertial range, as shown in Fig.\,\ref{fig:PSD}(a). The experimentally-determined power law exponent of $-1.7 \pm 0.2$ is consistent with the well-known $-5/3$ power-lay decay based on Kolmogorov's theory of Newtonian turbulence \cite{kolmogorov1941local} and Taylor's frozen flow hypothesis \cite{taylor1938spectrum}, and expresses the balance between turbulent energy gain from the mean flow and the viscous dissipation of turbulent kinetic energy. 
% We report details on the validation of Taylor's frozen flow hypothesis for our jets in Appendix \ref{app:Kolmogorov}.

%Based on Kolmogorov's theory of Newtonian turbulence \cite{kolmogorov1941local} and Taylor's frozen flow hypothesis \cite{taylor1938spectrum}, the PSD of the velocity fluctuations $E(f)$, also referred to as turbulent kinetic energy spectrum, follows the scaling relation $E(f) \sim f^{-5/3}$.  As shown in Fig.\,\ref{fig:PSD}a, we calculate the turbulent kinetic energy spectrum at fixed Eulerian location $x = 40D_h$ and at the centerline of the jet by measuring the velocity fluctuations time series $u_{cl}(t)$. The power spectrum shows a power-law decay of $-1.7 \pm 0.2$ in the inertial range which is consistent with the well-known $-5/3$ power-lay decay. Consistent with the previous experiments in Newtonian turbulence \cite{Kundu2016fluid}, this power-law deviates from $-1.7 \pm 0.2$ at $f\simeq 100$ which is smaller than the Kolmogorov frequency $f_k = 340 \textrm{ Hz}$. More details on the calculation of the Kolmogorov frequency and validation of Taylor's frozen flow hypothesis for the studied jet are shown in Appendix \ref{app:Kolmogorov}. 

\noindent The turbulent kinetic energy spectrum evaluated at the centerline of the viscoelastic jet with $\textrm{El} = 0.013$ also exhibits self-similar behavior but with a different power-law decay characterised by an exponent of $-2.9 \pm 0.3$, as shown in Fig.\,\ref{fig:PSD}(b). We have shown in our previous work \cite{yamani2021spectral} that the turbulent kinetic energy spectrum for viscoelastic jet flow contains a frequency range over which polymer chains sustain a persistent time-averaged rate of strain, $\langle\dot{\varepsilon}\rangle \sim U_{cl}/\delta$, through repeated extension and relaxation \cite{hinch1977mechanical}. Dimensional analysis then requires $E(k) \sim \langle\dot{\varepsilon}\rangle^2 k^{-3}$ \cite{vonlanthen2013grid}. Considering the frozen flow hypothesis \cite{taylor1938spectrum}, $E(f) \sim f^{-3}$, where $f = (k/2\pi) U_{cl}$, in agreement with our data over a frequency range referred to as the elasto-inertial frequency range of the spectrum. We have also determined the turbulent kinetic energy spectrum for the same viscoelastic jet at $\textrm{Re} = 400$ and $x/D_h = 30, 20$ and at $\textrm{Re} = 200$ and $x/D_h = 40, 30, 20$; all of the turbulent kinetic energy spectra exhibit a power-law with average slope $-2.8 \pm 0.5$ in the elasto-inertial range of frequencies.

\noindent The start of the inertial and elasto-inertial frequency range occurs approximately at the dimensionless frequency $f(\delta/U_{cl}) \sim 1$, as shown in Fig.\,\ref{fig:PSD}(a) and (b), denoting the condition at which the frequency and the associated wavenumber correspond to the time and length scales of the largest eddies that can be generated in the flow. We note that for $f(\delta/U_{cl}) \lesssim 1$ (for the largest eddies and the largest turbulent fluctuations), the PSD in the viscoelastic spectrum is higher than in the Newtonian spectrum. At higher frequencies, however, the PSD of the viscoelastic jet rapidly falls below the Newtonian one due to its steeper power-law decay, as shown in the inset of Fig.\,\ref{fig:PSD}(a). \textcolor{black}{The higher power in EIT compared to Newtonian turbulence at smaller frequencies, or equivalently for larger eddies, considering the frozen flow hypothesis \cite{taylor1938spectrum}, is consistent with the schlieren observation of larger LCSs in the streamwise direction for EIT compared to Newtonian turbulence, as shown in Fig.\,\ref{fig:SpatialResults}.} This is also an experimental evidence for the effect of polymer chains in altering the turbulent kinetic energy cascade by absorbing turbulence energy through their repeated extension and releasing it back to the flow through subsequent molecular relaxation \cite{dubief2004coherent,dubief2013mechanism,page_jfm2014,valente2014effect,valente2016energy,pereira2017active, Graham2014, xi2010active}. 

\noindent \textcolor{black}{
For Newtonian turbulence, dissipation becomes dominant at the Kolmogorov frequency where the smallest eddies are dissipated by viscosity. The Kolmogorov frequency, $f_k = 739 \textrm{ Hz}$ for our system is calculated in Appendix\,\ref{app:Kolmogorov}, and is marked with vertical dashed line in Fig.\,\ref{fig:PSD}(a) which  shows that the measurement of the turbulent kinetic energy spectrum from LDV cannot resolve the entire inertial range of the spectrum. This range of the spectrum can, however, be accessed by calculating the spatial PSD of the intensity fluctuations in schlieren images. We show in  Appendix\,\ref{app:psdSchlieren}   power-law decay rates for the spatial spectra (Fig.\,\ref{fig:PSDSpatial}(a) and (b)) that are consistent with the TKE frequency spectra shown in Fig.\,\ref{fig:PSD}(a) and (b) for Newtonian turbulence and EIT, respectively. In addition, the spatial PSD analysis extends the range of dimensionless frequency/wavenumber to two decades over which the $-5/3$ and $-3$ power-law decay rates are observed for Newtonian turbulence and EIT, respectively, as shown in Fig.\,\ref{fig:PSDSpatial}(a) and (b). Finally, the wavenumber associated with Kolmogorov length scale can be resolved and the deviation from the $-5/3$ power-law is observed, as shown in Fig.\,\ref{fig:PSDSpatial}(a).}

\noindent \textcolor{black}{EIT and its spectral properties have been areas of research interest and discovery.  Early theoretical work on EIT has reported that below a critical Reynolds number, where the product of polymer relaxation time and principal Lyapunov exponent (\textit{i.e.}, logarithmic rate of the divergence of two nearby Lagrangian trajectories) is smaller than unity, most polymers are in equilibrium and act as passive particles in the flow. Above this critical Reynolds number, however, the product of polymer relaxation time and principal Lyapunov exponent is larger than unity as a result of the polymers reaching an elongated state \cite{balkovsky2000turbulent, balkovsky2001turbulence} and contributing to the fluctuating flow by their energy storage and release. This contribution results in a power-law decay of the turbulent kinetic energy spectrum that is steeper than $-3$ \cite{fouxon2003spectra}. Reference \cite{vonlanthen2013grid}, however, discusses that this argument may not be conclusive as the physics changes when flow scales become smaller than the length of the polymer chains. The slope associated with the spectrum decay of EIT has also been studied in numerical simulations. While slopes as steep as $-14/3$ have been reported, without providing a physical or theoretical explanation \cite{dubief2013mechanism, thais2013spectral, watanabe2013hybrid}, it has been shown through theoretical arguments that when the polymer relaxation time is not significantly larger than the Kolmogorov timescale \cite{kolmogorov1941local}, the turbulent kinetic energy spectrum decays with a slope of $-3$ \cite{valente2014effect, valente2016energy}. For this range of relaxation and Kolmogorov time scales, a fraction of the turbulent energy is transferred to and stored in the polymer chains, preventing it from being transferred to small scales of motion through the conventional Newtonian turbulent energy cascade \cite{abreu2022turbulent}.  Turbulent channel flow experiments  \cite{vonlanthen2013grid, warholic1999influence, mitishita2022statistics}, and experiments of submerged round jets \cite{yamani2021spectral} all report a power-law decay with slope $-3$ for the turbulent kinetic energy spectrum of EIT. Direct numerical simulations of viscoelastic jets \cite{soligo2022non}, wakes \cite{guimaraes2022turbulent}, and Taylor-Couette flow \cite{lopez2022vortex} confirm slopes of $-3$. In this work, we calculate the PSD for Newtonian turbulence and EIT using velocity fluctuations and intensity fluctuations of schlieren images that are a surrogate for concentration fluctuations (\textit{cf.} Appendix\,\ref{app:psdSchlieren}). Using these two approaches, we  show a consistent $-3$ power-law decay for EIT over two decades of dimensionless frequency/wavenumber.}

\section{Conclusions}\label{sec:Conclusion}

\noindent Our experiments have revealed the emergence and evolution of inertio-elastic instabilities in submerged planar jets of dilute polymer solutions. We show that in jets characterized by small, non-zero elasticity a viscoelastic shear-layer instability emerges at the interface between the jet and the surrounding fluid that advances the transition to turbulence to lower Reynolds numbers and distances closer to the nozzle. This shear-layer instability evolves independently of a longer wavelength instability of the jet column. An increase in the elasticity number merges the sinuous shear-layer instability with the instability of the jet column. %The emergence of the shear layer instability destabilizes the jet while its merging with the jet column instability re-stabilizes the jet. 
%While linear stability analyses are not able to predict the nonlinear evolution of this shear layer instability, our experimental results show that the presence of the shear layer mode advances the transition to turbulence to lower Reynolds numbers and closer to the nozzle.
The two jet-column and shear-layer modes identified by our DMD analysis can be used to prescribe realistic initial perturbations in future simulations of turbulent viscoelastic jets. 

\noindent At high Reynolds numbers, the Newtonian and viscoelastic jets transition to two distinctly different states identified as Newtonian turbulence and EIT, respectively. Elasto-inertial turbulence is characterized by universal features that are observed in our experiments for a wide range of elasticity numbers: The rate of spreading of jets in the EIT state is independent of elasticity number, but slower than that observed for a Newtonian jet, which indicates a smaller rate of fluid entrainment for turbulent viscoelastic jets compared to Newtonian jets. Moreover, EIT exhibits Lagrangian coherent structures that are elongated in the streamwise direction with autocovariances that are independent of elasticity number. These structures are consistent with the strong elasticity-driven decay in the spectrum of EIT, which results in a higher spectral energy for larger flow structures as well as $f^{-3}$ power-law decay due to the contribution of the repeated stretching and relaxation of polymer chains to the turbulent energy cascade.

% Moreover, EIT exhibits a higher spectral energy for larger flow structures, independent of elasticity number, as well as a $f^{-3}$ power-law decay as a result of the contribution of the repeated stretching and relaxation of polymer chains to the turbulent energy cascade. This strong elasticity-driven decay in the spectrum and the high spectral energy of larger flow structures results in the viscoelastic jets exhibiting Lagrangian coherent structures that are elongated in the streamwise direction with autocovariances that are independent of elasticity number.

\noindent Our results motivate future numerical simulations of viscoelastic jets to probe the experimentally-observed universality of the EIT state for long chain flexible polymers over a wide range of elasticity numbers. Using data assimilation techniques, simulations that faithfully reproduce the measurements will also be able to provide predictive estimates of the full flow field in the jet, including the polymer conformation and elastic stresses.

\begin{acknowledgments}
\noindent We are grateful to Dr. James W. Bales for his advice on building the schlieren imaging setup. We thank MSE Inc. for providing the miniLDV G5B sensor. This work was supported by the National Science
Foundation (NSF) Grant No. CBET-2027870 to MIT and CBET-2027875 to JHU. We acknowledge the support of the Natural Sciences and Engineering Research Council of Canada (NSERC), funding reference number CGSD2-532512-2019. Cette recherche a été financée par le Conseil de recherches en sciences naturelles et en génie du Canada (CRSNG), numéro de référence CGSD2-532512-2019. 
\end{acknowledgments}

\appendix

\section{Calculation of power spectral density for LDV measurements}
\label{app:psd}

\noindent A time series measured with the LDV sensor is discrete and unevenly sampled over finite time, necessitating care in post-processing the data for spectral analysis. %If the LDV signal were evenly sampled and discrete over finite time, discrete Fourier transform (DFT) based algorithms such as FFT could be used to calculate the Fourier transform and the power spectral density of the signal. For unevenly sampled data, the first option is to lay down a grid of evenly spaced sampling times and interpolate values onto that grid. Interpolation, however, performs poorly in spectral analysis as it may cause spurious bulge of power at low frequencies due the large uneven gaps between sampled data \cite{press1992numerical}. An alternative method for calculating the power spectral density of unevenly sampled data was first proposed by Lomb \cite{lomb1976least} and later elaborated by Scargle \cite{scargle1982studies} and is commonly referred to as 
We employ the Lomb-Scragle periodogram method \cite{lomb1976least, scargle1982studies, press1992numerical}, assuming discrete values of $\tilde{U}_{cl}^i(x,t_i)$ are sampled at a fixed streamwise point $x$ and at times $t_i$, where $i = 1, \cdots, N$, such that there are $N$ sampled data points. Defining the  mean for the data,
\begin{equation}
    U_{cl}(x) = \frac{1}{N}\sum^N_{i=1} \tilde{U}_{cl}^i(x,t_i),
\end{equation}
% and 
% \begin{equation}
%     \sigma^2 = \frac{1}{N-1}\sum^N_{i=1} \left(\tilde{U}_{cl}^i(x,t_i) - U_{cl}(x)\right)^2.
% \end{equation}
the dimensional PSD with dimension $[\textrm{m}^2/\textrm{s}^2]$ is 
\begin{equation}
    E(f) = \frac{1}{N}\left[\frac{\left[\sum_{i=1}^N \left(\tilde{U}_{cl}^i(x,t_i) - U_{cl}(x)\right)\cos \omega \left(t_i - \tau \right)\right]^2}{\sum_{i=1}^N \cos^2 \omega \left(t_i - \tau \right)}+\frac{\left[\sum_{i=1}^N \left(\tilde{U}_{cl}^i(x,t_i) - U_{cl}(x)\right)\sin \omega \left(t_i - \tau \right)\right]^2}{\sum_{i=1}^N \sin^2 \omega \left(t_i - \tau \right)}\right],
\end{equation}
\noindent where $\omega = 2\pi f$ and $\tau$ is evaluated from 
\begin{equation}
    \tan \left(2\omega \tau \right) = \frac{\sum_{i=1}^N \sin 2\omega t_i}{\sum_{i=1}^N \cos 2\omega t_i}.
\end{equation}
% \noindent This method is equivalent to estimating the harmonic content of the data set by a linear least squares fitting of the form 
% \begin{equation}
%     \tilde{U}_{cl}(x,t) = A \cos \omega t + B \sin \omega t.
% \end{equation}
\noindent  %Lomb-Scragle periodogram has a significantly higher computational time compared to DFT based methods, such as FFT, due to the large number of trigonometric function evaluations for each frequency. Lomb-Scragle periodogram, however, weights data on a per point basis instead of a per time interval basis. A time interval basis weighting approach, such as DFT based methods, can cause a high error in the case of unevenly sampled data \cite{press1992numerical}.

The velocity time series, which is partly shown in Fig.\,\ref{fig:LDV}(a) for a Newtonian turbulent jet at $\textrm{Re} = 400$, is segmented into one second blocks with $50\%$ overlap. The PSD for each segment is calculated using the Lomb-Scragle periodogram method, and all segments are averaged (solid line in Fig.\,\ref{fig:LDV}(b)).  The frequencies are divided into logarithmically spaced bins and the mean power spectrum is averaged within each bin (hollow symbols shown in Fig.\,\ref{fig:LDV}(b)). %More details on segmenting and Lomb-Scragle periodogram can be found in chapter 13 of \cite{press1992numerical}.

\section{Evolution equation for the radius of elastic filament in CaBER based on FENE-P model} \label{app:FENE-P-CaBER}

\noindent We follow the derivation in \cite{wagner2015analytic} for the evolution equation of the filament radius over time based on the FENE-P model in the elasto-capillary regime. The dimensionless filament radius $\xi(t) = R(t)/R_0$ is given in implicit form by   

\begin{equation}
    -\frac{\left(b+3\right)^2}{b\left(b+2\right)}\tau = \left(\frac{1}{1+E_c(b+3)}-\frac{1}{1+\xi E_c(b+3)}\right) +3\ln \left(\frac{1+\xi E_c(b+3)}{1+E_c(b+3)}\right)+4E_c\frac{(b+3)}{(b+2)}\left(\xi-1\right),
\label{eq:FENE-P}
\end{equation}
where $b = 3L_{max}^2$, $\tau = t/\lambda_{\textrm{FENE-P}}$ is time nondimensionalized with the relaxation time parameter of the FENE-P model,  $E_c$ is the elasto-capillary number defined as $E_c = GR_0/\sigma$, the ratio between the elastic modulus of the suspension of dumbbells ($G \sim Nk_BT$) and the capillary pressure at the beginning of the elasto-capillary region  $\sigma/R_0$.
The prediction of equation \ref{eq:FENE-P} best matches the experimental data using an empirical numerical factor of 6 such that $G = 6Nk_BT$. See \cite{wagner2015analytic} for additional discussion of the early time response of the filament before an elasto-capillary balance is fully established. Equation\,\ref{eq:FENE-P} simplifies to the exponential form expected for the Oldroyd-B constitutive model (eq. \ref{eq:Oldroyd-B}) in the limit where $b \rightarrow \infty$. Both models predict the evolution accurately in the elasto-capillary region, as shown in Fig.\,\ref{fig:Rheology}(b); however, in the region where the finite extensibility of the polymer chains becomes important (later times), the FENE-P model (solid black line) provides a more accurate description of the data compared to the Oldroyd-B model (dashed black line).

\section{Flow visualization}\label{app:snapshots}

\begin{sidewaysfigure}[ht]
    \includegraphics[width=0.99\textwidth]{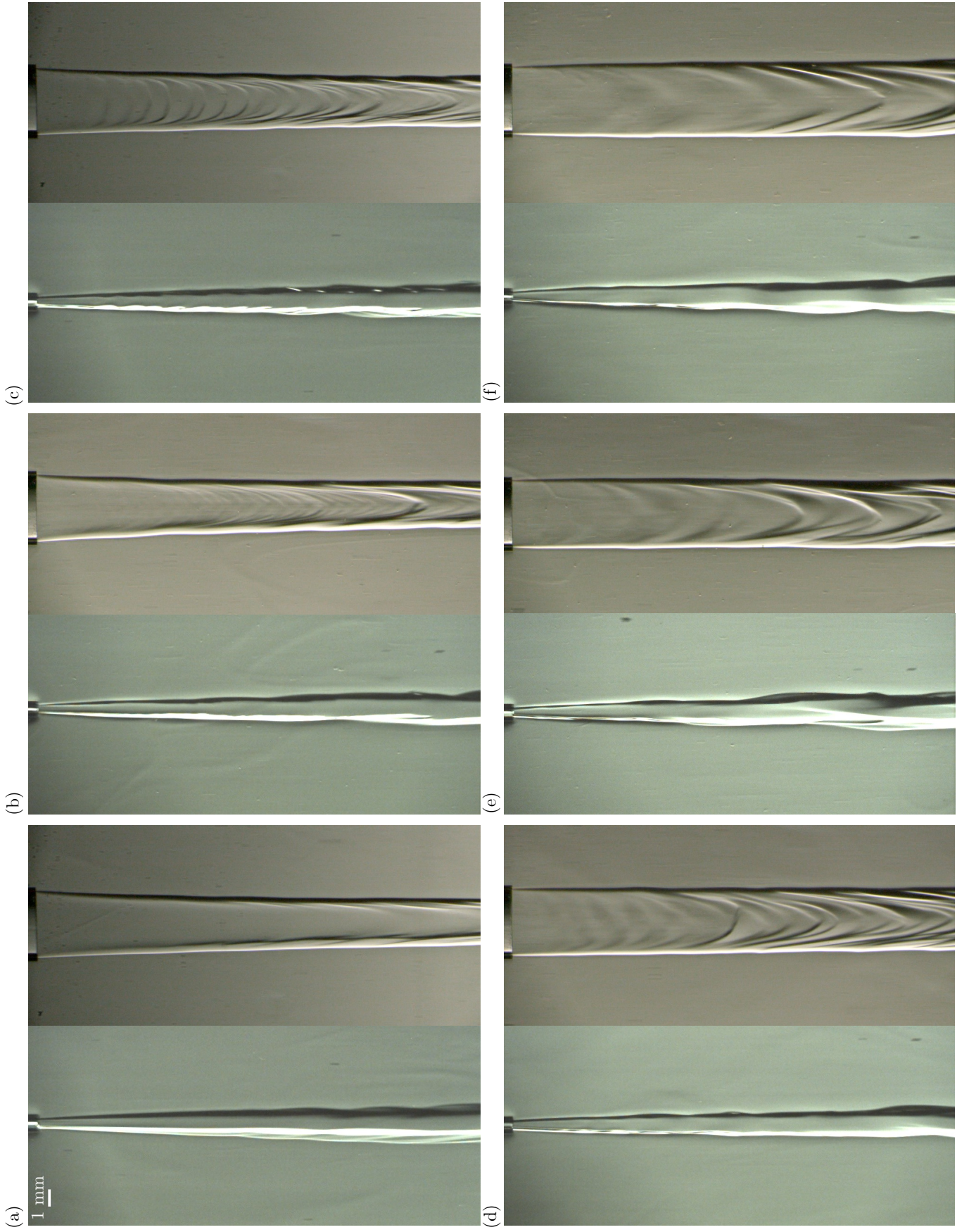}
    % \includesvg[width=0.99\textwidth]{ImagesFinal/Re100.svg}
    \caption{Side-view and front-view schlieren snapshots of (a) the Newtonian jet ($\textrm{El} = 0$), and viscoelastic jets with elasticity numbers of (b) $\textrm{El} = 0.004$, (c) $\textrm{El} = 0.013$, (d) $\textrm{El} = 0.041$, (e) $\textrm{El} = 0.066$, and (f) $\textrm{El} = 0.124$ at $\textrm{Re} = 100$. The elasticity number and hence the Weissenberg number, $\textrm{Wi} \sim \textrm{El} \cdot \textrm{Re}/(1-\beta)$, also increases from (a) to (f).}
    \label{fig:Re100AllSolutions}
\end{sidewaysfigure}

\begin{sidewaysfigure}[ht]
    \includegraphics[width=0.99\textwidth]{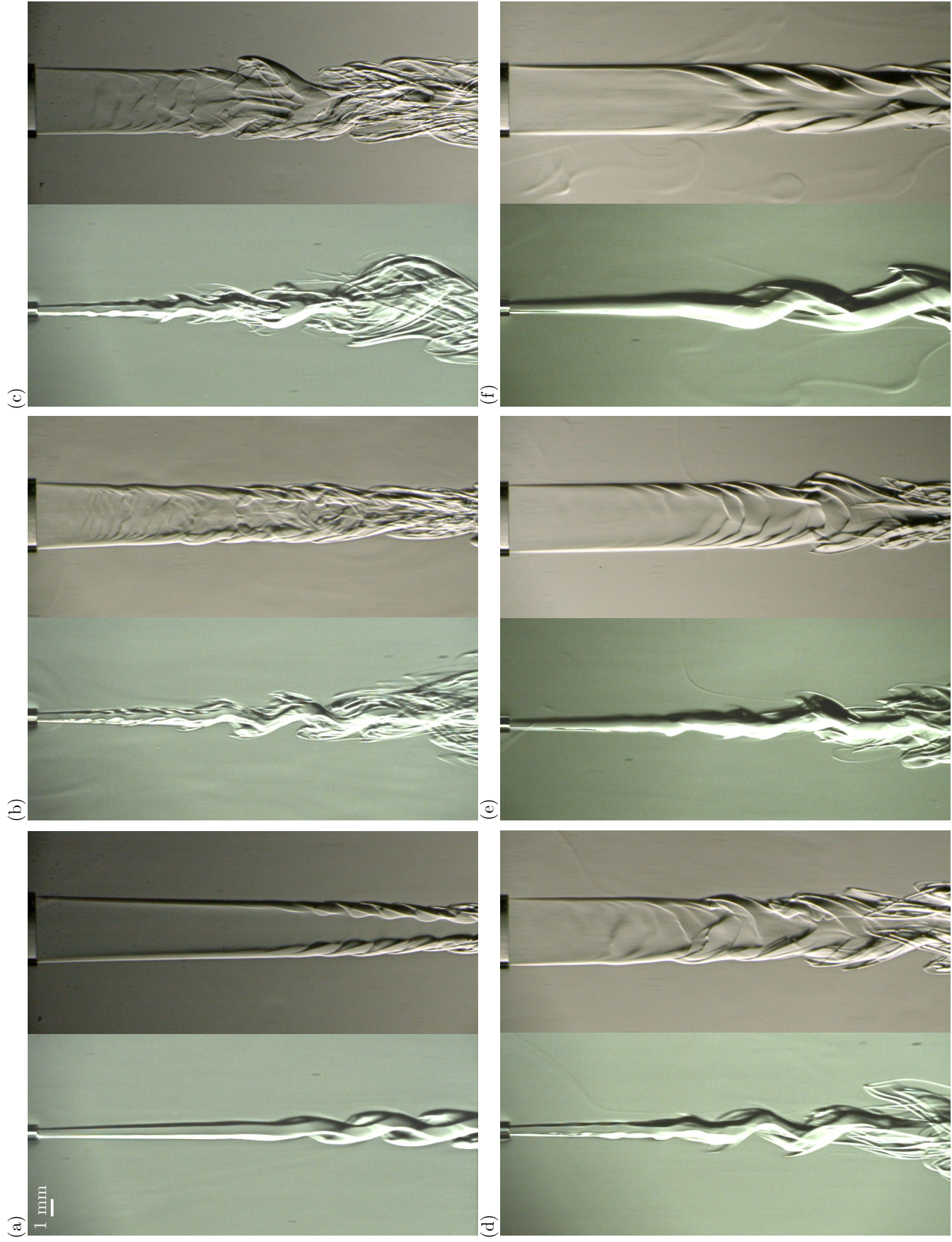}
    \caption{Side-view and front-view schlieren snapshots of (a) the Newtonian jet ($\textrm{El} = 0$), (b) $\textrm{El} = 0.004$, (c) $\textrm{El} = 0.013$, (d) $\textrm{El} = 0.041$, (e) $\textrm{El} = 0.066$, and (f) $\textrm{El} = 0.124$ at $\textrm{Re} = 200$. The elasticity number and hence the Weissenberg number also increases from (a) to (f).}
    \label{fig:Re200AllSolutions}
\end{sidewaysfigure}

\begin{sidewaysfigure}[ht]
    \includegraphics[width=0.99\textwidth]{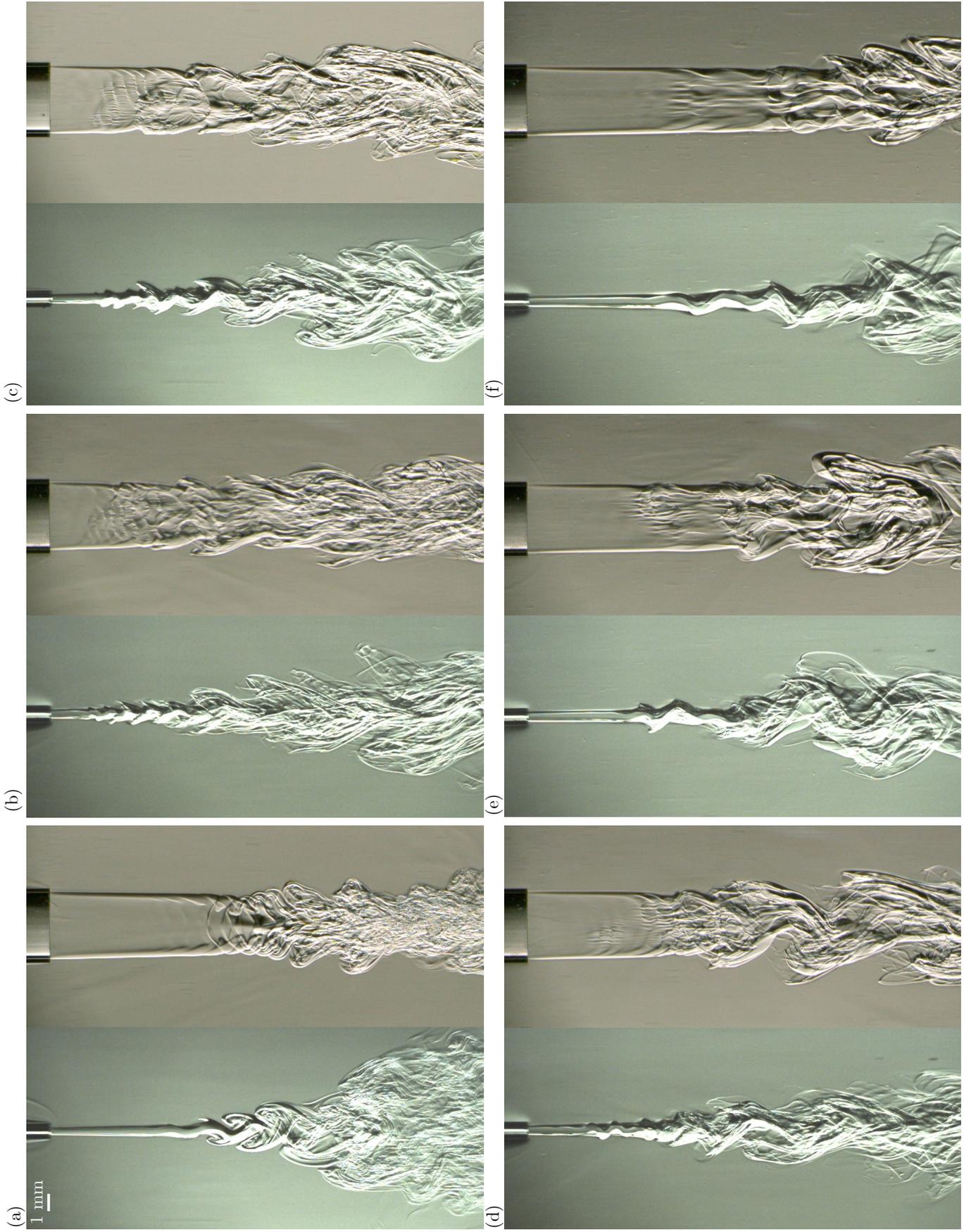}
    \caption{Side-view and front-view schlieren snapshots of  (a) the Newtonian jet ($\textrm{El} = 0$), (b) $\textrm{El} = 0.004$, (c) $\textrm{El} = 0.013$, (d) $\textrm{El} = 0.041$, (e) $\textrm{El} = 0.066$, and (f) $\textrm{El} = 0.124$ at $\textrm{Re} = 400$. The elasticity number and hence the Weissenberg number increases from (a) to (f).}
    \label{fig:Re400AllSolutions}
\end{sidewaysfigure}

\begin{sidewaysfigure}[ht]
    \includegraphics[width=0.99\textwidth]{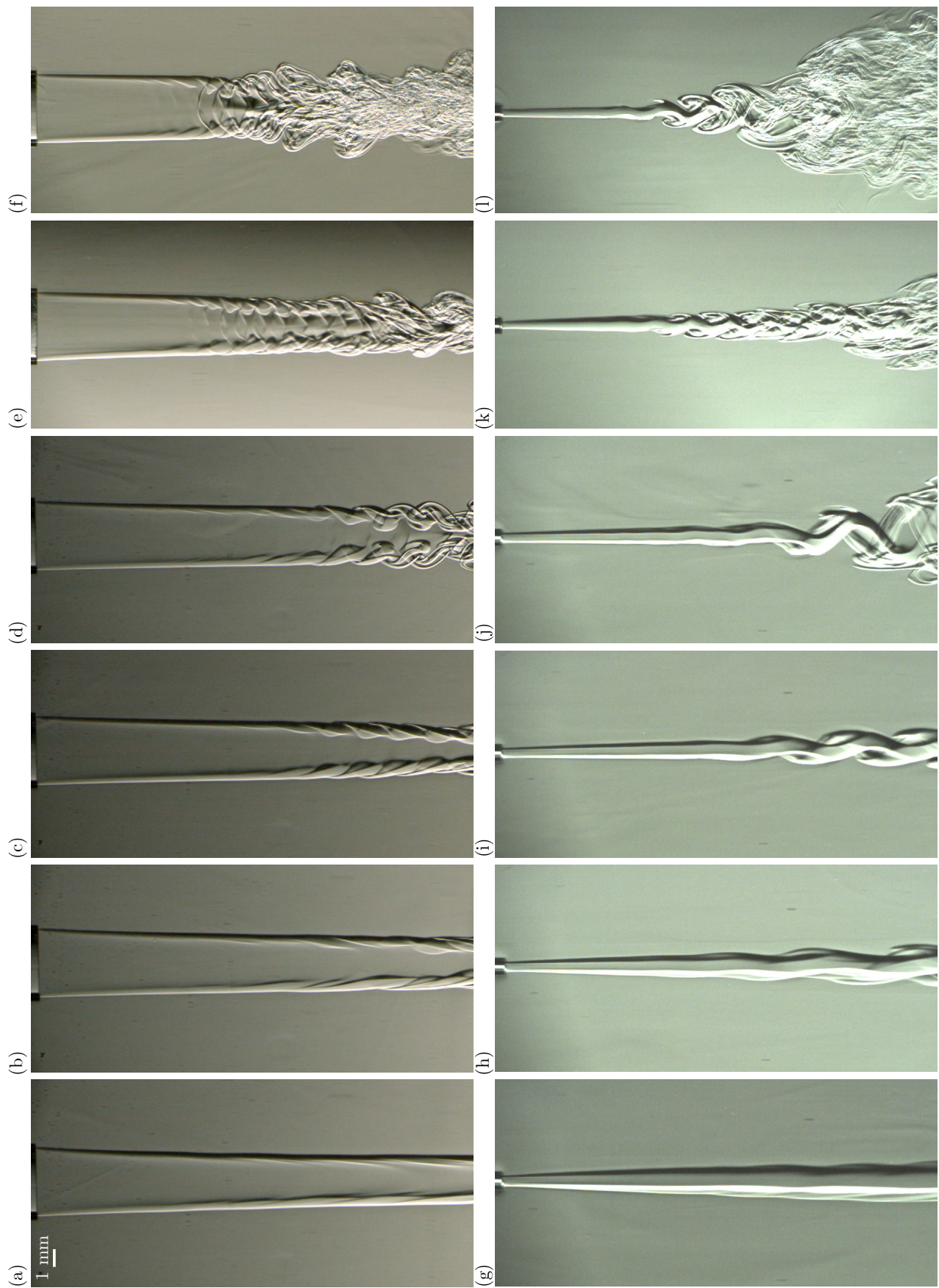}
    \caption{Front-view schlieren snapshots of the Newtonian jet ($\textrm{El} = 0$) at (a) $\textrm{Re} = 100$, (b) $\textrm{Re} = 150$, (c) $\textrm{Re} = 200$, (d) $\textrm{Re} = 250$,  (e) $\textrm{Re} = 300$, and (f) $\textrm{Re} = 400$. Side-view schlieren snapshots of the Newtonian jet ($\textrm{El} = 0$) at (g) $\textrm{Re} = 100$, (h) $\textrm{Re} = 150$, (i) $\textrm{Re} = 200$,  (j) $\textrm{Re} = 250$, (k) $\textrm{Re} = 300$, and (l) $\textrm{Re} = 400$.}
    \label{fig:NewtonianAllRe}
\end{sidewaysfigure}

\begin{sidewaysfigure}[ht]
    \includegraphics[width=0.99\textwidth]{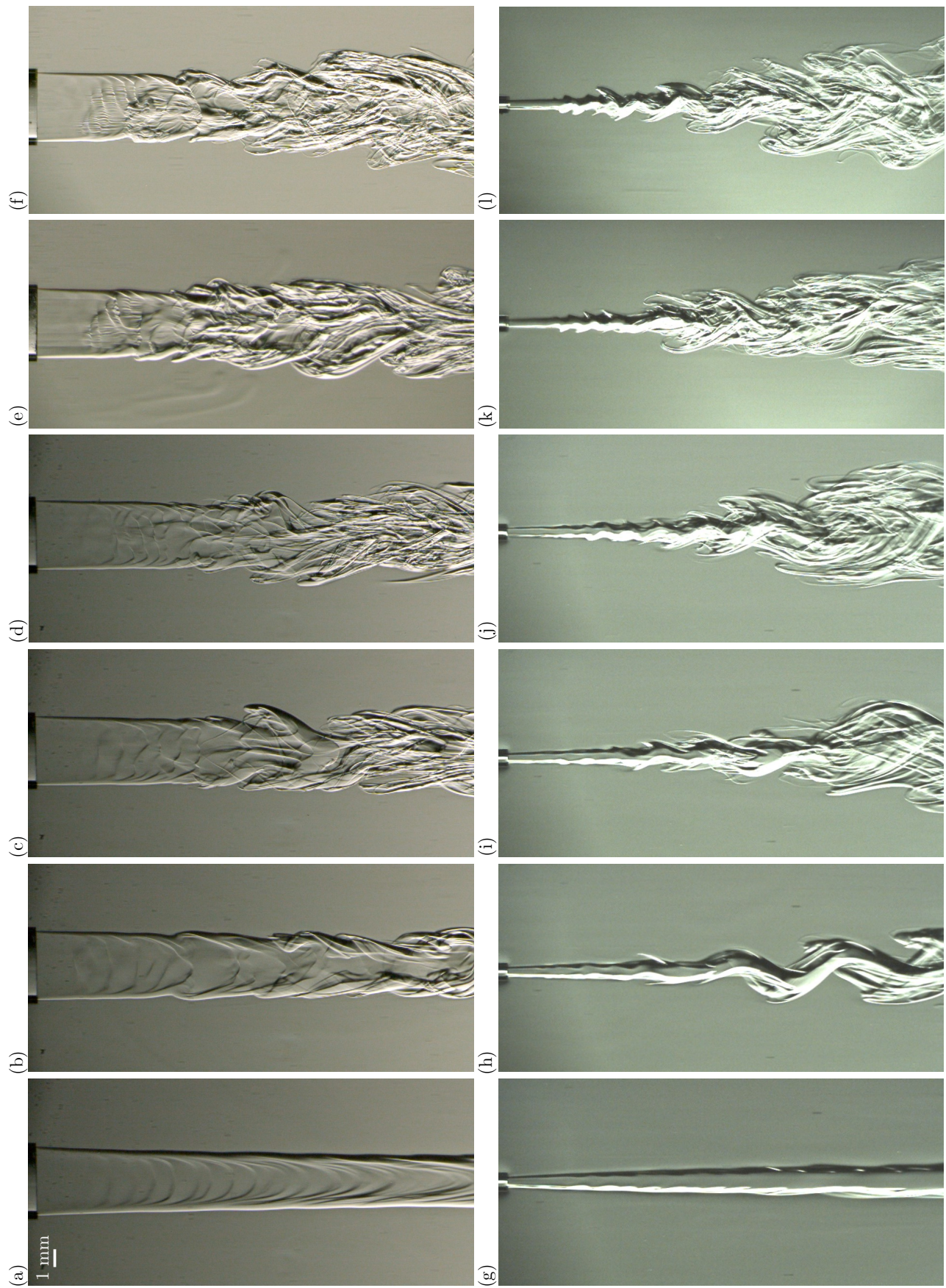}
    \caption{Front-view schlieren snapshots of a weakly viscoelastic jet ($\textrm{El} = 0.013$) at (a) $\textrm{Re} = 100$, (b) $\textrm{Re} = 150$, (c) $\textrm{Re} = 200$, (d) $\textrm{Re} = 250$,  (e) $\textrm{Re} = 300$, and (f) $\textrm{Re} = 400$. Side-view schlieren snapshots of the \textcolor{black}{same jet ($\textrm{El} = 0.013$) at} (g) $\textrm{Re} = 100$, (h) $\textrm{Re} = 150$, (i) $\textrm{Re} = 200$,  (j) $\textrm{Re} = 250$, (k) $\textrm{Re} = 300$, and (l) $\textrm{Re} = 400$.}
    \label{fig:PEO4M100ppmAllRe}
\end{sidewaysfigure}

\begin{sidewaysfigure}[ht]
    \includegraphics[width=0.99\textwidth]{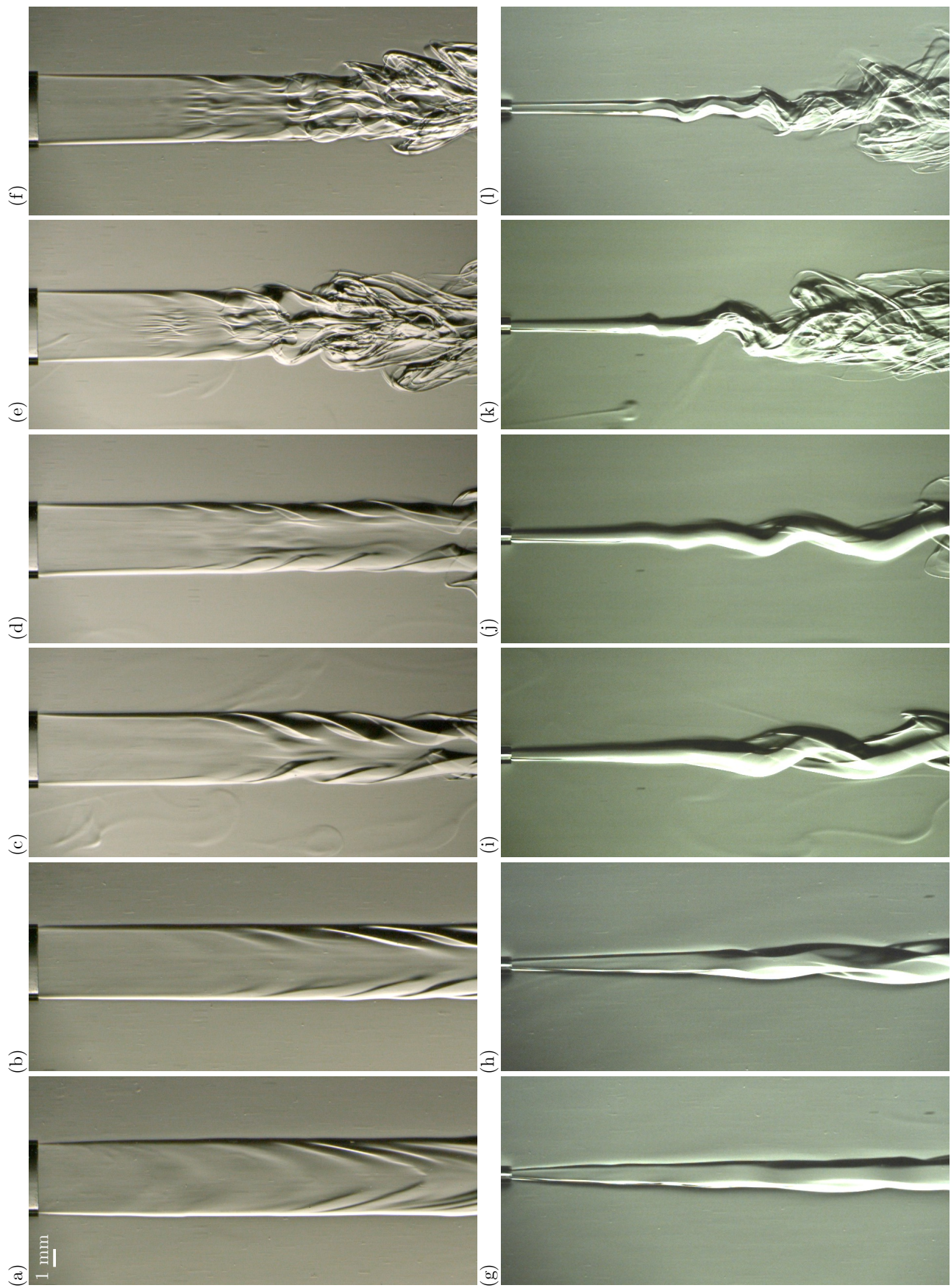}
    \caption{Front-view schlieren snapshots of a more viscoelastic jet ($\textrm{El} = 0.124$) at (a) $\textrm{Re} = 100$, (b) $\textrm{Re} = 150$, (c) $\textrm{Re} = 200$, (d) $\textrm{Re} = 250$,  (e) $\textrm{Re} = 300$, and (f) $\textrm{Re} = 400$. Side-view schlieren snapshots of \textcolor{black}{the same jet ($\textrm{El} = 0.124$) at} (g) $\textrm{Re} = 100$, (h) $\textrm{Re} = 150$, (i) $\textrm{Re} = 200$,  (j) $\textrm{Re} = 250$, (k) $\textrm{Re} = 300$, and (l) $\textrm{Re} = 400$.}
    \label{fig:PEO8M150ppmAllRe}
\end{sidewaysfigure}

\noindent Front-view and side-view snapshots of Newtonian and viscoelastic jets, with different elasticity numbers, are shown at different Reynolds numbers. %Images such as these are used for DMD computations and the analysis of the rate of jet spreading. 
Figures\,\ref{fig:Re100AllSolutions}-\ref{fig:Re400AllSolutions} show the impact of increasing elasticity number from $\textrm{El} = 0$ to $\textrm{El} = 0.124$ at fixed Reynolds numbers ranging from $\textrm{Re} = 100$ to $\textrm{Re} = 400$. Figures\,\ref{fig:NewtonianAllRe}-\ref{fig:PEO8M150ppmAllRe} show the impact of increasing Reynolds number from $\textrm{Re} = 100$ to $\textrm{Re} = 400$ at fixed \textcolor{black}{elasticity numbers} ranging from $\textrm{El} = 0$ to $\textrm{El} = 0.124$. 

% \begin{sidewaysfigure}[ht]
%     \includegraphics[width=0.99\textwidth]{images/Re150Elchangev2.png}
%     \caption{Side-view and Front-view Schlieren snapshots of the (a) Newtonian jet ($\textrm{El} = 0$), (b) $\textrm{El} = 0.004$, (c) $\textrm{El} = 0.013$, (d) $\textrm{El} = 0.041$, (e) $\textrm{El} = 0.066$, and (f) $\textrm{El} = 0.124$ at $\textrm{Re} = 150$. Elasticity number and hence Weissenberg number increases from (a) to (f).}
%     \label{fig:Re150AllSolutions}
% \end{sidewaysfigure}

% \begin{sidewaysfigure}[ht]
%     \includegraphics[width=0.99\textwidth]{images/Re250Elchangev2.png}
%     \caption{Side-view and Front-view Schlieren snapshots of the (a) Newtonian jet ($\textrm{El} = 0$), (b) $\textrm{El} = 0.004$, (c) $\textrm{El} = 0.013$, (d) $\textrm{El} = 0.041$, (e) $\textrm{El} = 0.066$, and (f) $\textrm{El} = 0.124$ at $\textrm{Re} = 250$. Elasticity number and hence Weissenberg number increases from (a) to (f).}
%     \label{fig:Re250AllSolutions}
% \end{sidewaysfigure}

% \begin{sidewaysfigure}[ht]
%     \includegraphics[width=0.99\textwidth]{images/Re300Elchangev2.png}
%     \caption{Side-view and Front-view Schlieren snapshots of the (a) Newtonian jet ($\textrm{El} = 0$), (b) $\textrm{El} = 0.004$, (c) $\textrm{El} = 0.013$, (d) $\textrm{El} = 0.041$, (e) $\textrm{El} = 0.066$, and (f) $\textrm{El} = 0.124$ at $\textrm{Re} = 300$. Elasticity number and hence Weissenberg number increases from (a) to (f).}
%     \label{fig:Re300AllSolutions}
% \end{sidewaysfigure}

% \begin{sidewaysfigure}[ht]
%     \includegraphics[width=0.99\textwidth]{images/PEO4M50ppmReincreasev2.png}
%     \caption{Front-view Schlieren snapshots of viscoelastic jet ($\textrm{El} = 0.004$) at (a) $\textrm{Re} = 100$, (b) $\textrm{Re} = 1500$, (c) $\textrm{Re} = 200$, (d) $\textrm{Re} = 250$,  (e) $\textrm{Re} = 300$, and (f) $\textrm{Re} = 400$. Side-view Schlieren snapshots of the Newtonian jet ($\textrm{El} = 0$) at (g)$\textrm{Re} = 100$, (h) $\textrm{Re} = 150$, (i) $\textrm{Re} = 200$,  (j) $\textrm{Re} = 250$, (k) $\textrm{Re} = 300$, and (l) $\textrm{Re} = 400$.}
%     \label{fig:PEO4M50ppmAllRe}
% \end{sidewaysfigure}

% \begin{sidewaysfigure}[ht]
%     \includegraphics[width=0.99\textwidth]{images/PEO4M200ppmReincreasev2.png}
%     \caption{Front-view Schlieren snapshots of viscoelastic jet ($\textrm{El} = 0.041$) at (a) $\textrm{Re} = 100$, (b) $\textrm{Re} = 1500$, (c) $\textrm{Re} = 200$, (d) $\textrm{Re} = 250$,  (e) $\textrm{Re} = 300$, and (f) $\textrm{Re} = 400$. Side-view Schlieren snapshots of the Newtonian jet ($\textrm{El} = 0$) at (g)$\textrm{Re} = 100$, (h) $\textrm{Re} = 150$, (i) $\textrm{Re} = 200$,  (j) $\textrm{Re} = 250$, (k) $\textrm{Re} = 300$, and (l) $\textrm{Re} = 400$.}
%     \label{fig:PEO4M200ppmAllRe}
% \end{sidewaysfigure}

% \begin{sidewaysfigure}[ht]
%     \includegraphics[width=0.99\textwidth]{images/PEO4M300ppmReincreasev2.png}
%     \caption{Front-view Schlieren snapshots of viscoelastic jet ($\textrm{El} = 0.066$) at (a) $\textrm{Re} = 100$, (b) $\textrm{Re} = 1500$, (c) $\textrm{Re} = 200$, (d) $\textrm{Re} = 250$,  (e) $\textrm{Re} = 300$, and (f) $\textrm{Re} = 400$. Side-view Schlieren snapshots of the Newtonian jet ($\textrm{El} = 0$) at (g)$\textrm{Re} = 100$, (h) $\textrm{Re} = 150$, (i) $\textrm{Re} = 200$,  (j) $\textrm{Re} = 250$, (k) $\textrm{Re} = 300$, and (l) $\textrm{Re} = 400$.}
%     \label{fig:PEO4M300ppmAllRe}
% \end{sidewaysfigure}

\newpage

\section{Dynamic mode decomposition (DMD)}\label{app:DMD}

\begin{figure*}[tbhp]
\centering
\includegraphics[width=0.99\textwidth]{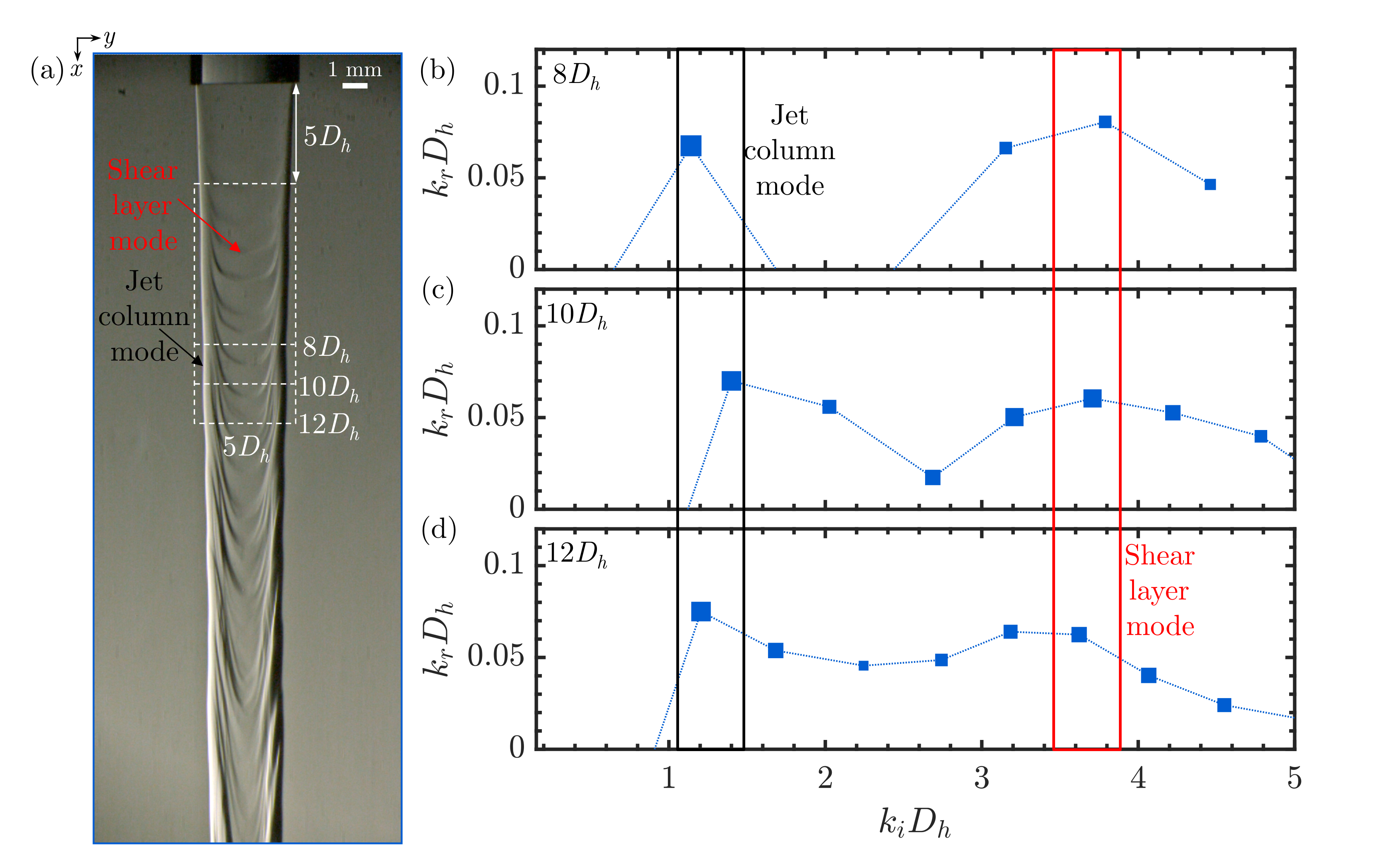}
\caption{(a) Front-view of the viscoelastic jet with $\textrm{El} = 0.013$ at $\textrm{Re} = 100$. The shear-layer and jet-column modes are identified with red and black arrows, respectively. The three investigated regions of interrogation are marked with dashed white lines. Wavenumber and growth rate of the jet for regions of interrogation with streamwise length (b) $8D_h$, (c) $10D_h$, and (d) $12D_h$. The abscissa and ordinate are nondimensionalized with the jet hydraulic diameter $D_h$. The size of the symbols denotes the relative magnitude of the amplitude of each mode $|b_j|$. The loci of the two peaks are marked with black and red bounding rectangles for the jet-column mode and shear-layer mode, respectively. The dashed lines are guides to the eye for identifying the local maxima in growth rate.}
\label{fig:DMDROI}
\end{figure*}

\noindent The DMD results for a viscoelastic jet with $\textrm{El} = 0.013$ and at $\textrm{Re} = 100$ using different regions of interrogation are shown in Fig.\,\ref{fig:DMDROI}. The regions of interrogation have the spanwise width $5D_h$ and varying streamwise lengths of $8D_h$, $10D_h$, and $12D_h$. Two wavenumber peaks at similar wavenumbers are observed for any choice of region of interrogation, confirming the robustness of the method used to identify the jet-column mode (marked by the  black vertical lines) and the shear-layer mode (marked by the red vertical lines), as shown in Fig.\,\ref{fig:DMDROI}(b)-(d). Note that smaller regions of interrogation restrict identification of low wavenumber modes, whereas larger regions of interrogation also potentially include the nonlinear evolution of the linear modes of the instability exhibited by the jet.

\begin{figure*}[tbhp]
\centering
\includegraphics[width=0.99\textwidth]{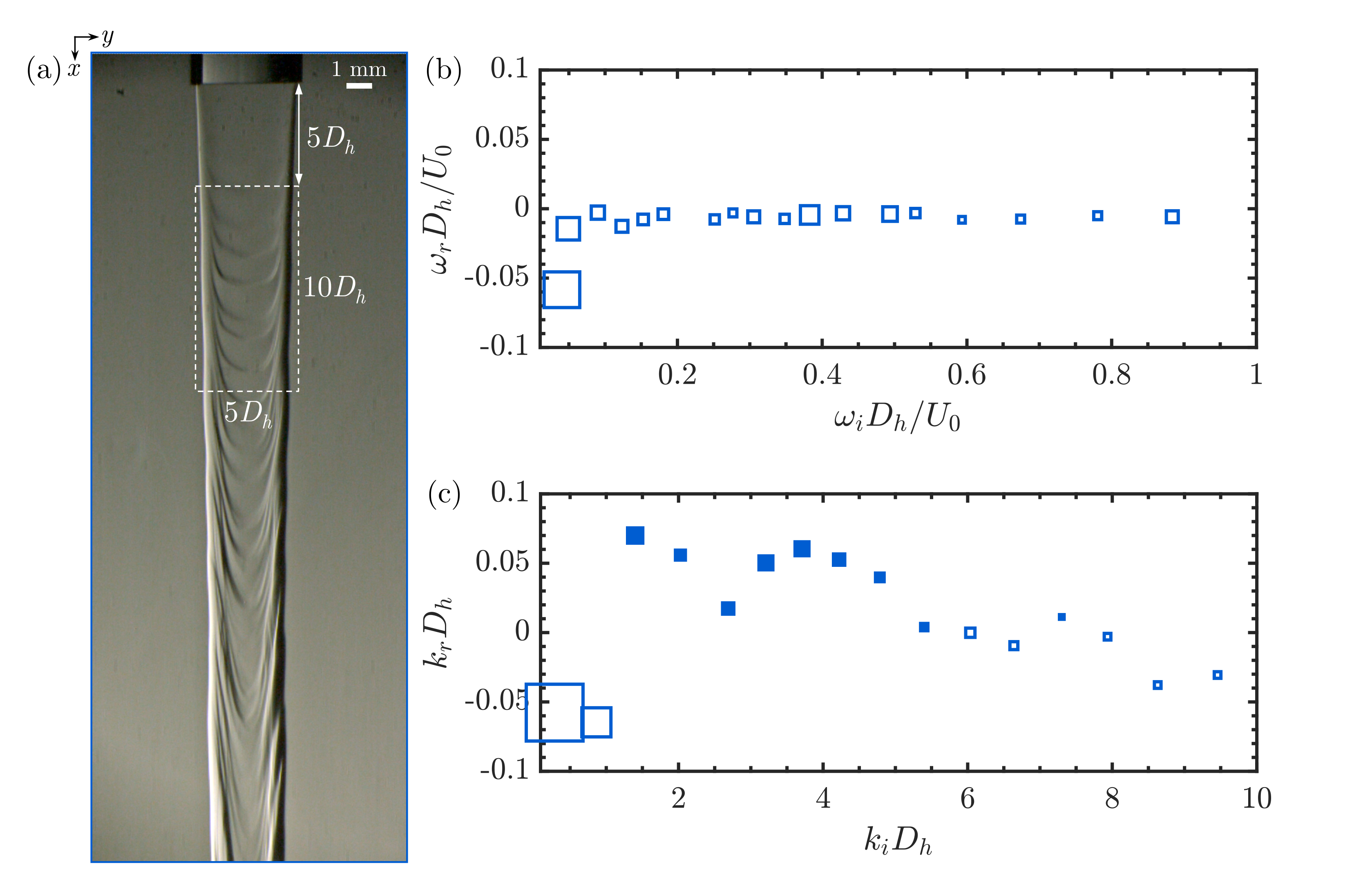}
\caption{(a) Front-view snapshot of a viscoelastic jet with $\textrm{El} = 0.013$ at $\textrm{Re} = 100$. (b) Frequency and temporal growth rate of the jet determined by temporal DMD analysis for the region of interrogation marked with white dashed lines in (a). The abscissa and ordinate are nondimensionalized with the jet hydraulic diameter $D_h$ and the mean velocity at the exit of the nozzle $U_0$. (c) Wavenumber and spatial growth rate of the jet determined by spatial DMD analysis. The abscissa and ordinate are nondimensionalized with the jet hydraulic diameter $D_h$. The size of the symbols denotes the relative magnitude of the amplitude of each mode $|b|$ with respect to the DC mode, \textit{i.e.}, $\omega_i = 0$ and $k_i = 0$ for the temporal and spatial DMD analyses, respectively. Filled and hollow symbols denote unstable and stable modes, respectively.}
\label{fig:DMDTemporalvsSpatial}
\end{figure*}

\noindent Similar to the spatial DMD discussion in Section\,\ref{sec:DMD}, the analysis can be performed temporally to decompose the disturbances visualized by schlieren imaging into a set of orthogonal modes. The intensity signal $I(x,y,t)$ is expressed as a linear superposition of orthogonal modes $\phi(x, y)$, each having a complex amplitude $b_j$ (where $|b_j|$ determines the relative importance of each mode), a frequency $\omega_i$ (which characterizes the periodicity of the mode in time, $t$) and a temporal growth rate $\omega_r$ (where $\omega_r>0$ represents a temporally growing mode):
\begin{equation}
    % I(x, y, t) - \bar{I} = \sum_{j=1}^{N_m} b_j \phi_j(y, t) \exp{(k_r^j + ik_i^j)x}, 
    I(x, y, t) = \sum_{j=1}^{N_t} b_j \phi_j(x, y) \exp{(\omega_r^j + i\omega_i^j)t}, 
\end{equation}
where $j=1,2,\,...\,, N_t$ and $N_t$ is the total number of modes. We use the same approach discussed in Section\,\ref{sec:DMD} to determine the retained number of modes for the temporal DMD analysis. Figure\,\ref{fig:DMDTemporalvsSpatial}(b) and (c) compare the results of the temporal and spatial DMD analyses. As expected, all modes determined in the temporal case are stable, \textit{i.e.}, $\omega_r<0$. The spatial analysis captures a number of downstream amplifying modes, $k_r>0$, indicating that the jet is spatially unstable and that the instability is convective in nature.

\section{Kolmogorov scale and the Taylor frozen field hypothesis}\label{app:Kolmogorov}

%%%%%%%%%%%%%uw expression%%%%%%%%%%%%%%%%%%%%%%%%%

\noindent \textcolor{black}{The production of turbulent kinetic energy scales as $\dot{W} \sim \overline{uw}\left(\partial U/\partial z \right)$ in the jet, where $u$ and $w$ are the streamwise and spanwise velocity fluctuations, $U$ is the mean streamwise velocity and $z$ is the spanwise direction. The largest eddies experience the mean shear, which is of order $U_{cl}/\delta$, where $U_{cl}(x)$ is the mean centerline velocity and $\delta(x)$ the thickness of the jet at streamwise location $x$ from the nozzle. Based on the maximum value reported for $\overline{uw}$ in the spanwise direction for experimental studies with Reynolds numbers that are at least an order of magnitude higher than our studied Reynolds numbers \cite{pope2000turbulent, viggiano2018turbulence, stanley2002study}, we estimate the order of magnitude for $\overline{uw}$ to be $ \sim 0.025 U_{cl}^2$ This yields $\dot{W} \propto 0.025 U_{cl}^2\left(U_{cl}/\delta\right)$. 
An equilibrium argument, while admittedly crude, can provide a useful estimate of the turbulent kinetic energy dissipation rate, $\bar{\epsilon} \sim \dot{W}$. The associated Kolmogorov length scale is defined as $\ell_{k} = \left(\nu^3/\bar{\epsilon}\right)^{1/4}$. For the Newtonian jet at streamwise distance $x = 40D_h$ from the nozzle, $U_{cl} = 0.1034 ~\textrm{m/s}$ and $\delta = 0.0103 ~\textrm{m}$, which yields $\bar{\epsilon} = 2.7 \times 10^{-3}$~m$^2$/s$^3$ and $\ell_{k} = 1.4 \times 10^{-4}$ m. Because of the weak  dependence of  $\ell_{k}$ on $\bar{\epsilon}$, changes in the coefficient of 
scaling for $\overline{uw}$, \textit{i.e.}, $0.025$, does not change the estimate for $\ell_{k}$ appreciably.  Using Taylor's frozen flow hypothesis \cite{taylor1938spectrum}, the Kolmogorov length scale calculated at the centerline of the Newtonian jet and at streamwise distance $x = 40D_h$ from the nozzle can be converted into a frequency scale, \textit{i.e.}, $f_k = (\ell_k/U_{cl})^{-1} = 739$ Hz, as shown in Fig.\,\ref{fig:PSD}(a).}

\section{Calculation of power spectral density for schlieren measurements}
\label{app:psdSchlieren}

\noindent \textcolor{black}{Schlieren images of the turbulent region of the jet provide spatial visualization of the changes in turbulent structures. The PSD of this signal can be calculated using FFT. 
We consider the spatial range $38D_h<x<42D_h$, and the centerline of the jet.  We denote the schlieren signal as $I^j(x_j,t)$ at fixed time $t$ and Eulerian streamwise position $x_j$, where $j = 1, \cdots, M$ and $M = 4D_h / \Delta = 104$ is the total number of pixels of size $\Delta$. 
\noindent \textcolor{black}{The PSD, $P(k)$, is normalized using the variance of the signal, $I^j(x_j,t)$. The results are time-averaged over 4000 frames, and the images are recorded at 4000 frames per second.}}

\noindent \textcolor{black}{
% The PSD evaluated from the schlieren images provides spatial information that complements the temporal LDV measurement.
The intensity of the schlieren images represents the local concentration fluctuations as the jet mixes with the background fluid in the turbulent region \cite{yamani2021spectral}. At high Schmidt number, $\textrm{Sc} = \nu/D$ (where $\nu$ is the kinematic viscosity of the dilute polymer solution and $D$ is the diffusivity of the individual polymer chains in the solvent), the PSD of concentration fluctuations exhibits an `inertio-convective range' with the same power-law decay rate as the inertial range of the turbulent kinetic energy spectrum \cite{oboukhov1949structure, corrsin1951spectrum}.  A requirement for this similarity is a high Schmidt number, which is satisfied in our Newtonian and viscoelastic solutions ($\sim 10^3$ and $\sim 10^6$, respectively \cite{yamani2021spectral}). }

\begin{figure*}[h]
\centering
\includegraphics[width=0.99\textwidth]{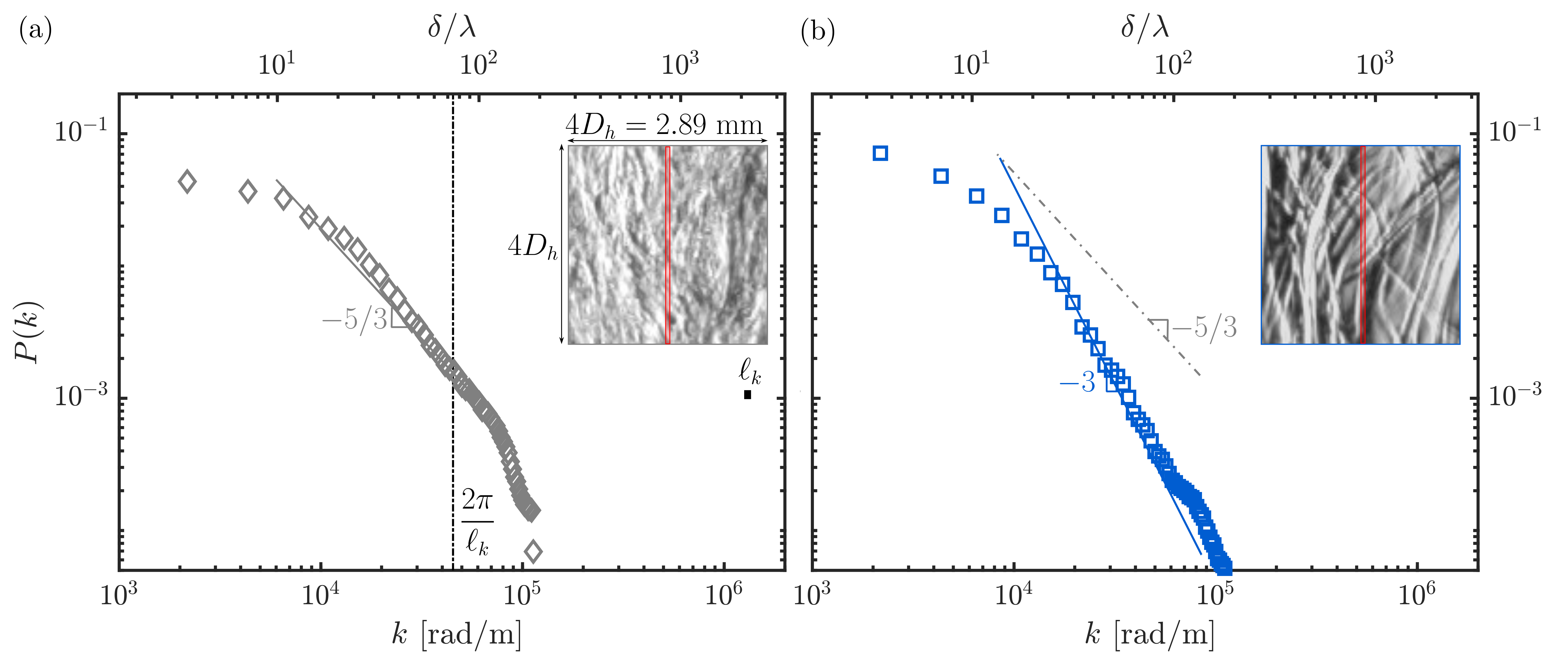}
\caption{\textcolor{black}{PSD of concentration fluctuations for (a) the Newtonian jet ($\MyDiamond[draw=water,fill=water]$) and (b) the viscoelastic jet (\mysquare{4M100ppm}) with $\textrm{El} = 0.013$ at $\textrm{Re} = 400$, calculated over a streamwise range $38D_h<x<42D_h$ from the nozzle and along the centerline of the turbulent jet. The upper axes are nondimensionalized with $\delta$ and $\lambda  = k/2\pi$. The insets of (a) and (b) show snapshots from schlieren videos used to calculate the PSD for the Newtonian and viscoelastic jets. The red rectangles mark the spatial signal used for each jet at the centerline of the jet. The Kolmogorov length scale ($\ell_k$) for the Newtonian jet, compared to the size of the spatial domain, is shown by the black line below the inset. The vertical dashed line in (a) marks the wavenumber associated with the Kolmogorov length scale.}}
% \vspace{-14pt}
\label{fig:PSDSpatial}
\end{figure*}

\noindent \textcolor{black}{Figures\,\ref{fig:PSDSpatial}(a) and (b) show the time-averaged and normalized PSD of the intensity, $P(k)$, as a function of streamwise wavenumber, for the Newtonian and viscoelastic ($\textrm{El} = 0.013$) jets. The insets show a localized region of size $4D_h$ along the centerline of the jets, for which the power spectral densities are calculated. The power-law decays for the Newtonian and viscoelastic jets are the same as the respective frequency spectra of the turbulent kinetic energy (\textit{cf.} Fig.\,\ref{fig:PSD}(a) and (b)).  To compare the scales spanned by the wavenumber spectra (Fig.\,\ref{fig:PSDSpatial}(a) and (b)) and the frequency spectra (Fig.\,\ref{fig:PSD}(a) and (b)), the upper abscissae are nondimensionalized with the jet spreading parameter $\delta$ (defined in Sec.\,\ref{sec:similarity}) and $\delta/U_{cl}$, respectively.  The latter captures the energy cascade for large to medium size eddies, while the wavenumber spectra extend to small eddies.  The Kolmogorov wavenumber is marked by a vertical dashed line in Fig.\,\ref{fig:PSDSpatial}(a) at $k = 2\pi/\ell_k$, where $\ell_k$ is calculated in Appendix\,\ref{app:Kolmogorov}.  For the Newtonian jet, this scale is consistent with the end of the $-5/3$ power-law decay.  The change in the spectral decay at this scale is less pronounced for the  viscoelastic jet (Fig.\,\ref{fig:PSDSpatial}(b)), at least for the range of wavenumbers accessible from the schlieren images.}  

\newpage
\bibliography{references.bib}

\end{document}